\documentclass[12pt,english]{article}
\usepackage[T1]{fontenc}
\usepackage[utf8]{inputenc}
\usepackage{geometry}
\usepackage{xcolor}
\usepackage{float}
\usepackage{textcomp}
\usepackage{amsmath}
\usepackage{amssymb}
\usepackage{graphicx}
\usepackage{setspace}
\usepackage[authoryear]{natbib}
\PassOptionsToPackage{normalem}{ulem}
\usepackage{ulem}
\usepackage{multibib}
\newcites{SIcite}{References}

\usepackage[colorlinks=true, allcolors=blue]{hyperref}

\makeatletter

\providecommand{\tabularnewline}{\\}
\providecolor{lyxadded}{rgb}{0,0,1}
\providecolor{lyxdeleted}{rgb}{1,0,0}

\DeclareRobustCommand{\lyxsout}[1]{\ifx\\#1\else\sout{#1}\fi}

\newcommand{\lyxaddress}[1]{
	\par {\raggedright #1
	\vspace{1.4em}
	\noindent\par}
}

\@ifundefined{date}{}{\date{}}
\usepackage{xcolor}

\newcommand{\bx}{\boldsymbol{x}}
\newcommand{\bxi}{\boldsymbol{\xi} }
\usepackage[normalem]{ulem}

\usepackage{lineno}
\usepackage{authblk}
\usepackage{ae,lmodern}

\renewenvironment{abstract}
 {\small
  \begin{center}
  \bfseries \abstractname\vspace{-.5em}\vspace{0pt}
  \end{center}
  \list{}{
    \setlength{\leftmargin}{.5cm}%
    \setlength{\rightmargin}{\leftmargin}%
  }%
  \item\relax}
 {\endlist}

\DeclareMathOperator\erf{erf}

\newcommand{\ellB}{\ell_{\mathrm{B},i}} 

\usepackage{babel}
\usepackage{url}

\makeatother

\begin{document}

\title{Local intraspecific aggregation in phytoplankton model communities:
spatial scales of occurrence and implications for coexistence}
\author{Coralie Picoche\textonesuperior $^{,*}$ , William R. Young\texttwosuperior ,
Fr\'e{}d\'e{}ric Barraquand\textonesuperior $^{,*}${}}
\maketitle
\thispagestyle{empty}
\vspace{-1cm}
\lyxaddress{\noindent \begin{center}
\textonesuperior Institute of Mathematics of Bordeaux, University
of Bordeaux and CNRS, Talence, France\\
 \texttwosuperior Scripps Institution of Oceanography, La Jolla, California,
USA 
\par\end{center}}
\onehalfspacing
\vspace{-0.5cm}
\begin{abstract}
The coexistence of multiple phytoplankton species despite their reliance
on similar resources is often explained with mean-field models assuming
mixed populations. In reality, observations of phytoplankton indicate
spatial aggregation at all scales, including at the scale of a few
individuals. Local spatial aggregation can hinder competitive exclusion
since individuals then interact mostly with other individuals of their
own species, rather than competitors from different species. To evaluate
how microscale spatial aggregation might explain phytoplankton diversity
maintenance, an individual-based, multispecies representation of cells
in a hydrodynamic environment is required. We formulate a three-dimensional
and multispecies individual-based model of phytoplankton population
dynamics at the Kolmogorov scale. The model is studied through both
simulations and the derivation of spatial moment equations, in connection
with point process theory. The spatial moment equations show a good
match between theory and simulations. We parameterized the model based
on phytoplankters' ecological and physical characteristics, for both
large and small phytoplankton. Defining a zone of potential interactions
as the overlap between nutrient depletion volumes, we show that local
species composition---within the range of possible interactions---depends
on the size class of phytoplankton. In small phytoplankton, individuals remain in mostly monospecific clusters. Spatial structure therefore favours intra- over inter-specific interactions for small phytoplankton, contributing to coexistence. Large phytoplankton cell neighbourhoods appear more mixed. Although some small-scale self-organizing spatial structure remains and could influence coexistence mechanisms, other factors may need to be explored to explain diversity maintenance in large phytoplankton.
\end{abstract}


$^{*}$correspondence to: \verb|frederic.barraquand@u-bordeaux.fr|
\& \verb|cpicoche@gmail.com|

Published in \textit{Journal of Mathematical Biology} with \verb|DOI:10.1007/s00285-024-02067-y|

\clearpage{}

\section*{Introduction}

Phytoplankton communities are among the most important photosynthetic
groups on Earth, being at the bottom of the marine food chain, and
responsible for approximately half the global primary production \citep{field_primary_1998}.
Their contribution to ecosystem functions is only matched by their
contribution to biodiversity. Indeed, phytoplankton communities are
characterized by a surprisingly high number of species. For example,
a single sample as small as a few mL can contain up to seventy species
\citep{REPHY_db,widdicombe_2021}. This observation is usually called
the ``paradox of the plankton'' \citep[a term coined by][]{hutchinson_paradox_1961}, which refers to the conflict between the observed diversity of species competing for similar resources in a seemingly homogeneous environment, and models predicting that only a few species will persist by outcompeting the others \citep{macarthur_competition_1964,huisman_biodiversity_1999,schippers_does_2001}.
Phytoplankton models for coexistence are now almost as diverse as
their model organisms \citep{record_paradox_2014}, but they often
describe only a handful of species, which does not correspond to the
diversity observed in the field. When modeling rich communities (\textgreater{}
10 species), classical answers to the plankton paradox involving temporal
fluctuations \citep[e.g.,][]{li_effects_2016,chesson_updates_2018}
are not sufficient to maintain a realistic diversity. For instance,
we found that a phytoplankton community dynamics model with environmental
fluctuations and storage effect still requires extra niche differentiation
for coexistence, which manifests in stronger intraspecific than interspecific
interactions \citep{picoche_how_2019}. However, it is not clear that
we should resort to hidden niches to explain phytoplankton coexistence,
as most models also make hidden simplifying assumptions that could
be relaxed. One that we relax here is mean-field dynamics at the microscale.
Indeed, field observations have revealed phytoplankton patchiness
for decades, with early records in the past centuries \citep{bainbridge_size_1957,stocker_marine_2012},
from the macro- to the micro-scale \citep{leonard_interannual_2001,doubell_high-resolution_2006,font-munoz_advection_2017}.

Phytoplankton patchiness can at least be partly explained by the hydrodynamics
of their environment: the size of these organisms is mostly below
the size of the smallest eddy (i.e., the Kolmogorov scale). In a typical
aquatic environment such as the ocean, phytoplankton individuals are
embedded in viscous micro-structures \citep{peters_effects_2000}
while phytoplankton populations are displaced by a turbulent flow
at slighly larger scales \citep{martin_phytoplankton_2003,prairie_biophysical_2012}.
Phytoplankton organisms therefore live in an environment where fluid
viscosity dominates at the scale of an individual but turbulent dispersion
dominates on length scales characteristic of a small population of
those individuals \citep{estrada_effects_1987,prairie_biophysical_2012}.

This leads us to consider demography in the context of this environmental
variation created by hydrodynamic processes. Individual-based models
provide a convenient depiction of population dynamics and movement
at the microscale \citep{hellweger_bunch_2009}. In this framework,
population growth is a result of individual births and deaths. Aggregation
of individuals can emerge from local reproduction coupled with limited
dispersal, which can happen in a fluid where turbulence and diffusion
are not strong enough to disperse kin aggregates \citep{young_reproductive_2001}.
The resulting local aggregation can then affect the community dynamics
at larger spatial scales, even when all competitors are equivalent
(i.e., with equal interaction strengths irrespective of species identity).
Indeed, the combination of local dispersal after reproduction and
local interactions leads to stronger intraspecific interactions than
interspecific interactions at the population level \citep{detto_stabilization_2016}.
This mechanism stabilizes the community, as a high intra-to-interspecific
interaction strength ratio makes a species control its abundance more
than it controls the abundance of other species, which is associated
with coexistence in theoretical models \citep{levine_importance_2009,barabas_self-regulation_2017}
and often observed in the field at the population level \citep{adler_competition_2018,picoche_strong_2020}.
Therefore, the microscale spatial distribution of individuals likely
affects the interaction structure within a community,
and may sustain diversity \citep{haegeman_how_2008}.

Existing models of phytoplankton populations near the Kolmogorov scale
--- between 1 mm and 1 cm in an oceanic environment \citep{barton_impact_2014}
--- focus on a single species and the clustering of its individuals
\citep{young_reproductive_2001,birch_master_2006,bouderbala_3d_2018,breier_emergence_2018}. These models share similarities to dynamic point process models \citep{law_population_2003,bolker_spatial_1999,plank_spatial_2015}
developed initially with larger organisms in mind. When phytoplankton
individual-based models consider multiple types of organisms, they
focus for now on how organisms with opposite characteristics \citep[e.g., increase versus decrease in density with turbulence in][]{borgnino_turbulence_2019,arrieta_fate_2020}
segregate spatially, or on coexistence of species that have contrasting
trait values \citep[e.g., size in][]{benczik_coexistence_2006}. Such models are useful as an explanation of how species with marked differences might coexist. The difficulty of the coexistence problem, however, is that we also have to explain how closely related species or genera (e.g., within diatoms), many of whom have similar size, buoyancy, chemical composition, etc., manage to coexist within a single trophic level.
This requires modelling \emph{similar} species in a spatially realistic
environment and objectively quantifying whether they aggregate or
segregate in space.

To do so, we build a multispecies version of the Brownian Bug Model
(BBM) of \citet{young_reproductive_2001}, an individual-based model
which includes an advection process mimicking a turbulent fluid flow,
passive diffusion of organisms, as well as stochastic birth and death
processes. The initial version of this model \citep{young_reproductive_2001}
coupled limited dispersal and local reproduction with ocean-like microscale
hydrodynamics, and showed spatial clusters of individuals of the same
species. The original BBM was limited to a single species and was
illustrated with two-dimensional simulations. The model was not strongly
quantitative \citep{picoche_rescience_2022} in the sense that parameters
were not informed by current knowledge on phytoplankton biology (numbers
of cells per liter, diffusion characteristics, etc.). As phytoplankton
organisms live in a three-dimensional environment, informing the model
with more realistic parameters requires us to shift to three dimensions.
We also extend the model to multiple species, and consider two size
classes for our phytoplankton communities, which are either made of
nanophytoplankton (3~\textmu m diameter, $\approx10^{6}$~cells~L$^{-1}$)
or microphytoplankton (50~\textmu m, $\approx10^{4}$~cells~L$^{-1}$).
We populate each community with 3 to 10 different species.

The Brownian Bug Model (in its original single-species form as in
the multispecies version considered here) is related to spatial branching
processes. Without advection, it combines a continuous-time, discrete-state
model for population growth and a continuous-time, continuous-space
Brownian motion for particle diffusion \citep{birch_master_2006}.
It is further complexified by a turbulent flow in \citet{young_reproductive_2001,picoche_rescience_2022}
as well as here. In spite of this complexity, it remains possible
to derive the dynamics of pair density functions, which quantify the
degree of intra- and interspecific clustering of organisms, via correlations
between positions of organisms (see next section). Thus we can understand
emergent spatial structures in analytic detail and compare these predictions to the results from three-dimensional simulations. Furthermore, because we do not consider direct interactions between organisms, the multispecies spatial point process that represents the stable state of the BBM is a random superposition of spatial point processes for each species
\citep{illian2008statistical}. This enables us to derive, in addition
to pair correlation functions, analytical formulas for the species
composition in the neighbourhood of an individual, which are more
readily ecologically interpreted than pair density or correlation
functions.

\section*{Model and spatial statistics}

\subsection*{Brownian Bug Model}

The Brownian Bug Model (BBM) describes the dynamics of individuals
in a turbulent and viscous environment, including demographic processes.
The model is continuous in space and time. Here we extend the mostly
two-dimensional, monospecific version in \citet{young_reproductive_2001},
to three dimensions and $S$ species.

Each individual is characterized by its species identity $i$ and
its position $\mathbf{x}^{T}=(x,\,y,\,z)$. The population dynamics
are modelled by a linear birth-death process with birth rate $\lambda_{i}$
and death rate $\mu_{i}$. Each individual independently follows a
Brownian motion with diffusivity $D_{i}$, and is advected by a common
stochastic and chaotic flow modelling turbulence. The model applies
in the Batchelor regime, which means that the separation $s(t)$ between
two individuals $k$ and $l$ grows exponentially with time with stretching
parameter $\gamma$, i.e., $s(t)=\ln\left(|\bx_{k}-\bx_{l}|(t)\right)\propto3\gamma t$
\citep{kraichnan_convection_1974,young_reproductive_2001}.

Within a given community (the set of all individuals of the $S$ species),
all species share the same parameters: $\lambda_{i}$, $\mu_{i}$
and $D_{i}$ values can change between communities, as we later consider
small and large phytoplankton, but are set to common values within
a community. On the contrary, $\gamma$ describes the environment
and is not community-specific, i.e., all individuals are displaced
by the same turbulent stirring. For numerical simulations, time needs
to be discretized (this is required for diffusion and advection modelling).
The approximated model advances through time in small steps of duration
$\tau$. During each interval, events unroll as follows:
\begin{enumerate}
\item Demography: each individual can either reproduce with probability
$p_{i}=\lambda_{i}\tau$ (forming a new individual of the same species
$i$ at the same position $\mathbf{x}$ as the parent), die with probability
$q_{i}=\mu_{i}\tau$, or remain unchanged with probability $1-p_{i}-q_{i}$.
\item Diffusion: each individual moves to a new position $\mathbf{x}(t')=\mathbf{x}(t)+\delta\mathbf{x}(t)$,
with $t<t'<t+\tau$. The random displacement $\delta\mathbf{x}(t)$
is drawn from a Gaussian distribution $\mathcal{N}(0,\Delta_{i}^{2})$
with $D_{i}=\Delta_{i}^{2}/2\tau$ the diffusivity. This diffusive
step separates the initially coincident pairs produced by reproduction
in step 1 above.
\item Turbulence: each individual is displaced by a turbulent flow, modelled
with the Pierrehumbert map \citep{pierrehumbert_tracer_1994}, adapted
to three dimensions following \citet{ngan_scalar_2011}. Thus given
the position at time $t'$ the updated position at time $t+\tau$
is
\end{enumerate}
\begin{equation}
\begin{array}{cc}
x(t+\tau) & =\qquad x(t')+\frac{U\tau}{3}\cos\left(ky(t')+\phi(t)\right)\\
y(t+\tau) & =\qquad y(t')+\frac{U\tau}{3}\cos\left(kz(t')+\theta(t)\right)\\
z(t+\tau) & =z(t')+\frac{U\tau}{3}\cos\left(kx(t+\tau)+\psi(t)\right).
\end{array}
\end{equation}
Above, $U$ is the velocity of the flow, $k=2\pi/L_{s}$ is the wavenumber
for the flow at the length scale $L_{s}$ (see below) and $\phi(t)$,
$\theta(t)$, $\psi(t)$ are random phases drawn from a uniform distribution
between $0$ and $2\pi$; these phases remain constant during the
interval between $t$ and $t+\tau$. The shift from continuous to
discrete-time turbulence modelling is described in Section~\ref{section:si_derivation} in the
Supplementary Information. The velocity $U$ is related to $\gamma$.
As the separation between two points grows exponentially with parameter
$3\gamma$ due to turbulence, the exponent $\gamma$ can be estimated
as the slope of $1/3\left\langle \ln(s(t))\right\rangle =f(t)$ in
the absence of diffusion and demography \citep{young_reproductive_2001,picoche_rescience_2022}.

Individuals are distributed in a cube of side length $L$, with periodic
boundary conditions. The cube dimensions are determined to balance
computing costs and realistic concentrations of individuals; they
represent the accumulation of a few volumes of scale $L_{s}$.

\subsection*{Characterization of the spatial distribution}

Let $W$ be the observation window (in our case, the whole cube, which
we never subsample hereafter). The state of the system at time $t$
can be described as a collection of $S$ populations, where the population
of species $i$ is made of $n_{i}$ individuals randomly distributed
in $W$, with positions $\boldsymbol{X}_{i}(t)=[\bx_{1,i}(t),\bx_{2,i}(t),\ldots,\bx_{n_{i},i}(t)]$.
$\boldsymbol{X}(t)~=~[\boldsymbol{X}_{1}(t),\ldots,\boldsymbol{X}_{S}(t)]$
arises from a stochastic and spatial individual-based model changing
through time, but can also be analyzed as a spatial point process
at time $t$. We note that the point distributions remain the same
for all spatial translations $\bxi$ (i.e., the point process described
by the set $\boldsymbol{X}~=~[\bx_{1},\bx_{2},\ldots,\bx_{k}]$ is the same
as $\boldsymbol{X_{\xi}}~=~[\bx_{1}+\bxi,\bx_{2}+\bxi,\ldots,\bx_{k}+\bxi])$:
the process is stationary.

A useful method to characterize a spatial point process is the use
of spatial moments (illustrated in Section~\ref{section:si_simple_point_processes} of the SI for simple
spatial point processes). These can be theoretically derived and used
to check simulations. The spatial moments of a process are, however,
merely statistical indicators which then need to be related to more
easily ecologically interpretable quantities. This is the role of
the dominance index, which we present below.

\subsubsection*{Spatial moments}

The first-order moment is the intensity of the process, or mean concentration
of individuals, whose empirical estimate is $C_{i}=\frac{\widehat{N_{i}(W)}}{V(W)}$,
where $\widehat{N_{i}(W)}$ is the empirical number of individuals
of species $i$ in the cube $W$ and $V(W)=L^{3}$ is the volume of
the cube; it does not give any information regarding the spatial distribution
of individuals, and their spatial correlations.

The second-order product density, or pair density $G(r,t)$, is the
expected density of pairs of points separated by a distance $r$ \citep{law_population_2003}.
A similar statistic can be used for marked spatial point process. In our case, the marks are the species' identities, and we can define $G_{ij}(r,t)$, so that $G_{ij}(r,t)d\mathbf{x}_{A}d\mathbf{x}_{B}$
is the probability of finding an individual of species $i$ in volume
$d\mathbf{x}_{A}$ and an individual of species $j$ in volume $d\mathbf{x}_{B}$, with the distance between the centers of $d\mathbf{x}_{A}$ and $d\mathbf{x}_{B}$ equal to $r$ (pages 219 and 325 in \citealp{illian2008statistical}).
We define $\boldsymbol{\xi}$ as the vector connecting the center
of $d\mathbf{x}_{A}$ to the center of $d\mathbf{x}_{B}$, while $r=|\boldsymbol{\xi}|$
is the radial distance. We show in \citet{picoche_rescience_2022}
that the intraspecific pair density $G_{ii}(r,t)$, in three dimensions,
is a solution of 
\begin{equation}
\frac{\partial G_{ii}}{\partial t}(r,t)=\frac{2D_{i}}{r^{2}}\frac{\partial}{\partial r}\left(r^{2}\frac{\partial G_{ii}}{\partial r}\right)+\frac{\gamma}{r^{2}}\frac{\partial}{\partial r}\left(r^{4}\frac{\partial G_{ii}}{\partial r}\right)+2(\lambda_{i}-\mu_{i})G_{ii}+2\lambda_{i}C_{i}\delta(\boldsymbol{\xi}).\label{eq:Young3D}
\end{equation}
The pair correlation function $g_{ij}(r,t)$, or pcf, can be derived
from the pair density and is defined as 
\begin{equation}
g_{ij}(r,t)=\frac{G_{ij}(r,t)}{C_{i}C_{j}}.\label{eq:def_pcf}
\end{equation}
The pcf is equal to one when the spatial distribution of species $i$
individuals is random relative to species $j$ individuals. To compute
the intraspecific pcf $g_{ii}(r,t)$ at steady state, considering
a population at equilibrium, we integrate Eq. \ref{eq:Young3D} (see
Appendices, Eqs. \ref{eq:steady_state}-\ref{eq:pcf_adv_bbm}) with
$\lambda_{i}=\mu_{i}$ and obtain 
\begin{equation}
g_{ii}(r)=1+\frac{\lambda_{i}}{4\pi D_{i}C_{i}\ellB}\left(\frac{\ellB}{r}+\arctan\left(\frac{r}{\ellB}\right)-\frac{\pi}{2}\right),\label{eq:pcf_intra_adv}
\end{equation}

where $\ellB=\sqrt{2D_{i}/\gamma}$ approximates the Batchelor scale
for species $i$.

The system converges rapidly to the solution in Eq. \ref{eq:pcf_intra_adv}
in the presence of advection. However, when there is no turbulent
advection, convergence is much slower, to the point that an equilibrium
assumption requires unrealistically long timeframes (see Section~\ref{section:si_convergence}
in the SI). We therefore need a time-dependent formula for the pcf
in the absence of advection, which can be obtained in the case where
$\gamma=0$ using a Green's function (see derivation in the Appendices,
Eqs. \ref{eq:G_no_adv}-\ref{eq:pcf_noadv_bbm}), 
\begin{equation}
g_{ii}(r,t)=1+\frac{\lambda_{i}}{4\pi rD_{i}C_{i}}\left\{ 1-\erf\left(\frac{r}{\sqrt{8D_{i}t}}\right)\right\} .\label{eq:pcf_intra_noadv}
\end{equation}
The above equations match when $\gamma\rightarrow0$ and $t\rightarrow+\infty$.

As populations of different species do not directly interact, each
population is an independent realization of a point process, which
means that the distribution of all individuals within the community
at time $t$ is a random superposition of stationary point processes
and thus $g_{ij}(r,t)~=~1$ if $i\neq j$ \citep[ p. 326, eq. 5.3.13]{illian2008statistical}.\medskip{}

Related to the pair correlation function is Ripley's $K$-function
$K(r)$. Using its marked version, $C_{j}K_{ij}(r)$ is the average
number of points of species $j$ surrounding an individual of species
$i$ within a sphere of radius $r$ \citep{illian2008statistical},
i.e., 
\begin{equation}
\forall r\geq0\textmd{, }K_{ij}(r)=\frac{1}{C_{j}}\mathbb{E}_{i}\left(N_{j}\left(b(o,r)\backslash\{o\}\right)\right),
\end{equation}
where $\mathbb{E}_{i}$ is the expectation with respect to individuals
of species $i$ and $N_{j}\left(b(o,r)\backslash\{o\}\right)$ is
the number of individuals of species $j$ in the sphere of radius
$r$ centered on individual $o$, not counting individual $o$ itself. $K_{ij}(r)$ is related to $g_{ij}(r)$ as 
\begin{equation}
g_{ij}(r)=\frac{K_{ij}'(r)}{4\pi r^{2}}.\label{eq:pcf_to_K}
\end{equation}
Combining Eq. \ref{eq:pcf_to_K} and, when $U>0$, Eq. \ref{eq:pcf_intra_adv},
we can show that (see Appendices, Eqs. \ref{eq:start-K}-\ref{eq:K_intra_adv_app})
\begin{equation}
K_{ii}(r)=\frac{4}{3}\pi r^{3}+\frac{\lambda_{i}r^{3}}{3D_{i}C_{i}\ellB}\left(\frac{\ellB}{r}+\frac{\ellB^{3}\log\left(\frac{r^{2}}{\ellB^{2}}+1\right)}{2r^{3}}+\arctan\left(\frac{r}{\ellB}\right)-\frac{\pi}{2}\right).\label{eq:K_intra_adv}
\end{equation}

When $U=0$, we need a time-dependent solution corresponding to our
simulation duration, i.e. (see Appendices, Eq. \ref{eq:start_K_intra_no_adv}-\ref{eq:end_K})
\begin{equation}
K_{ii}(r,t)=\frac{4}{3}\pi r^{3}+\frac{\lambda_{i}r^{2}}{C_{i}D_{i}}\left(\frac{1}{2}-\frac{1}{2}\erf\left(\frac{r}{\sqrt{8D_{i}t}}\right)\left(1-\frac{4D_{i}}{r^{2}}t\right)-\frac{\sqrt{2D_{i}t}}{\sqrt{\pi}r}e^{-\frac{r^{2}}{8D_{i}t}}\right).\label{eq:K_intra_no_adv}
\end{equation}

For random superposition of stationary point processes, $K_{ij}(r,t)=\frac{4}{3}\pi r^{3}$
if $i\neq j$ \citep[p. 324, eq. 5.3.5]{illian2008statistical}.

\subsubsection*{Dominance index}

The dominance index \citep[defined in Table S1 in the Supporting Information of][]{wiegand_how_2007}
is the ratio between the number of conspecifics and the number of
individuals of all species surrounding a given individual.

Let $M_{ij}(r)$ be the average number of individuals of species $j$
within a circle of radius $r$ around an individual of species $i$,
which can also be written with Ripley's $K$-function as $M_{ij}(r)=C_{j}K_{ij}(r)$.
$M_{ii}(r)$ corresponds to the conspecific neighbourhood and $M_{io}(r)=\sum_{j=1,j\neq i}^{S}M_{ij}(r)$
corresponds to individuals of all other species. We can then define
$\mathcal{D}_{i}$ as 
\begin{equation}
\begin{array}{ccc}
\mathcal{D}_{i}(r) & = & \frac{M_{ii}(r)}{M_{ii}(r)+M_{io}(r)}\\
 & = & \frac{C_{i}K_{ii}(r)}{\sum_{j=1}^{S}C_{j}K_{ij}(r)}.
\end{array}\label{eq:dominance}
\end{equation}
 When individuals of the same species $i$ tend to cluster, $\mathcal{D}_{i}(r)$
tends to 1 while it tends to the proportion of individuals of species
$i$ in the whole community when the distribution is uniform (Section~\ref{section:si_simple_point_processes} of the SI).

Using Eqs. \ref{eq:K_intra_adv} and \ref{eq:dominance}, we obtain
the formula for the dominance index in the presence of advection as
\begin{equation}
\mathcal{D}_{i}(r)=\frac{\frac{\lambda_{i}}{3D_{i}\ellB}\left(\frac{\ellB}{r}+\frac{\ellB^{3}\log\left(\frac{r^{2}}{\ellB^{2}}+1\right)}{2r^{3}}+\arctan\left(\frac{r}{\ellB}\right)-\frac{\pi}{2}\right)+\frac{4}{3}\pi C_{i}}{\frac{\lambda_{i}}{3D_{i}\ellB}\left(\frac{\ellB}{r}+\frac{\ellB^{3}\log\left(\frac{r^{2}}{\ellB^{2}}+1\right)}{2r^{3}}+\arctan\left(\frac{r}{\ellB}\right)-\frac{\pi}{2}\right)+\sum_{j=1}^{S}\frac{4}{3}\pi C_{j}}.\label{eq:dom_adv}
\end{equation}

In the absence of advection ($U=0,\gamma=0$), we use the time-dependent
dominance index, computed similarly: 
\begin{equation}
\mathcal{D}_{i}(r,t)=\frac{\frac{\lambda_{i}}{D_{i}r}\left(\frac{1}{2}-\frac{1}{2}\erf\left(\frac{r}{\sqrt{8D_{i}t}}\right)\left(1-\frac{4D_{i}}{r^{2}}t\right)-\frac{\sqrt{2D_{i}t}}{\sqrt{\pi}r}e^{-\frac{r^{2}}{8D_{i}t}}\right)+\frac{4}{3}\pi C_{i}}{\frac{\lambda_{i}}{D_{i}r}\left(\frac{1}{2}-\frac{1}{2}\erf\left(\frac{r}{\sqrt{8D_{i}t}}\right)\left(1-\frac{4D_{i}}{r^{2}}t\right)-\frac{\sqrt{2D_{i}t}}{\sqrt{\pi}r}e^{-\frac{r^{2}}{8D_{i}t}}\right)+\sum_{j=1}^{S}\frac{4}{3}\pi C_{j}}.\label{eq:dom_noadv}
\end{equation}

\subsection*{Parameters}

We model two types of organisms: microphytoplankton (defined by a
diameter between 20 and 200 \textmu m, here 50 \textmu m) and nanophytoplankton
(defined by a diameter between 2 and 20 \textmu m, here 3 \textmu m).
These two groups are characterized respectively by a low diffusivity,
slow growth and lower concentration vs. high diffusivity, fast growth
and higher concentration. Organisms are displaced by a turbulent fluid
whose velocity defines the time scale of the discretized model: we
give here the reasoning behind parameter values, keeping in mind that
our model parameters are only approximate. Main parameter definitions
and values are given in Table \ref{tab:Definition-and-value}.

\subsubsection*{Advection}

We first consider the advection process, due to the turbulence of
the environment. We only consider the Batchelor-Kolmogorov regime,
i.e., $L_S$ is below the size of the smallest
eddy, but above the smallest length scale of fluctuations in nutrient
concentrations. The defining scale of the environment therefore corresponds
to a Reynolds number 
\begin{equation}
\text{Re}=\frac{U}{k\nu}\approx1\label{eq:reynolds_def}
\end{equation}
where $\nu=10^{-6}$ m$^{2}$~s$^{-1}$ is the kinematic viscosity
for water. The smallest wavenumber $k$ corresponds to the largest
length scale $L_{s}$ (Kolmogorov scale), i.e., $k=2\pi/L_{s},$ with
$L_{s}\approx1$ cm in the ocean \citep{barton_impact_2014}. The
definition of the Reynolds number leads to 
\begin{equation}
\begin{array}{cc}
1 & \approx\;\frac{UL_{s}}{2\pi\nu}\\
\Leftrightarrow U & \approx\;\frac{2\pi\nu}{L_{s}}.
\end{array}\label{eq:compute_U}
\end{equation}

This means that $U=6.3\times10^{-4}$ m~s$^{-1}$ = $5.4\times10^{3}$
cm~d$^{-1}$. Using $U\tau/3=0.5$ cm as in \citet{young_reproductive_2001},
we have $\tau=2.8\times10^{-4}$ d $=24$ s. When $U\tau/3=0$, the
environment is only diffusive, we keep the same value for $\tau$.
For $U\tau/3=0.5$ cm, we estimate $\gamma=1231$ d$^{-1}$.

\subsubsection*{Diffusion}

If we use the Stokes-Einstein equations (\citealp{einstein1905molekularkinetischen},
cited from \citealp{dusenbery_2009}), diffusivity can be computed
with 
\begin{equation}
D_{i}=\frac{RT}{N_{A}}\frac{1}{6\pi\eta a_{i}}
\end{equation}
where $R=8.314$ J~K$^{-1}$~mol$^{-1}$ is the molar gas constant,
$T=293$ K is the temperature, $N_{A}=6.0225\times10^{23}$ is Avogadro's
number, $\eta=10^{-3}$ m$^{-1}~$kg~s$^{-1}$ is the dynamic viscosity
of water and $a_{i}$ is the radius of the organism considered.

Using $D_{i}=\frac{\Delta_{i}^{2}}{2\tau}$, we find that 
\begin{equation}
\begin{array}{cccc}
 & \Delta_{i} & = & \sqrt{2\tau D_{i}}\\
\Leftrightarrow & \Delta_{i} & = & \sqrt{\frac{RT}{N_{A}}\frac{\tau}{3\pi\eta a_{i}}}.
\end{array}
\end{equation}

We consider $a_{n}=1.5$ \textmu m for nanophytoplankton individuals
and $a_{m}=25$ \textmu m for microphytoplankton individuals, which
allows us to compute $\Delta_{n}$ and $\Delta_{m}$ (see Table \ref{tab:Definition-and-value}).

\subsubsection*{Ecological processes}

We study the community at equilibrium, with the birth rate equal to
the death rate, i.e., $p_{i}=q_{i}\,\forall i$. We use a microphytoplankton
doubling rate of $1$ d$^{-1}$ \citep{bissinger_predicting_2008}
and consider the fastest-growing nanophytoplankton species, corresponding
to a diameter of 3 \textmu m \citep{bec_growth_2008}, for which the
doubling rate is between 2 and 3 d$^{-1}$ (set to 2.5 d$^{-1}$ here).

\begin{table}[H]
\begin{centering}
\begin{tabular}{|c|c|c|}
\hline 
Parameter  & Definition  & Value\tabularnewline
\hline 
$p_{m},q_{m}$  & Probability of reproducing/dying for microphytoplankton & $2.8\times10^{-4}$\tabularnewline
$p_{n},q_{n}$  & Probability of reproducing/dying for nanophytoplankton & $6.9\times10^{-4}$\tabularnewline
$U$  & Turbulent advection speed  & \{0, 0.06\} cm.s$^{-1}$\tabularnewline
$\Delta_{m}$  & Diffusion parameter for microphytoplankton & $6.4\times10^{-5}$ cm\tabularnewline
$\Delta_{n}$  & Diffusion parameter for nanophytoplankton & $2.6\times10^{-4}$ cm\tabularnewline
\hline 
\end{tabular}
\par\end{centering}
\caption{Definitions and values of the main parameters used in the three-dimensional BBM, assuming the duration of a time step $\tau$ is 24 seconds. \label{tab:Definition-and-value}}
\end{table}

\subsubsection*{Range of interaction}

As we examine individual aggregation and its potential effects on
interactions between species, we have to ascertain the volume in which
an individual can be affected by the presence of other individuals,
or affect other individuals. We only consider here interactions due
to competition for nutrients, and therefore need to define a nutrient
depletion volume. We approximate this volume as the sphere of radius
$r$ where $C(r)\leq90\%C_{\infty}$ with $C_{\infty}$ the background
concentration of the nutrient and $C(a_i) = 0$ (perfect absorption at the cell surface). The radius of this nutrient depletion
volume is maximized when the individual is in stagnant water so that
diffusion is the only hydrodynamic process. In this case, the depletion
radius corresponds to 10 times the radius of the individual \citep{jumars_physical_1993,karp-boss_nutrient_1996}.
We define the maximum distance which allows for potential interactions
(due to competition for resources) between two individuals of radius
$a_{i}$ and $a_{j}$ as $d_{\text{threshold}}$, and the corresponding
volume of potential interactions around an organism as $V_{\text{int}}=4/3\pi d_{\text{threshold}}^{3}$
with 
\begin{equation}
d_{\text{threshold}}=10a_{i}+10a_{j}.\label{eq:distance_interaction}
\end{equation}
We consider this maximum value as our baseline, keeping in mind that
turbulence reduces the size of the nutrient depletion volume and increases
the nutrient flux to the cell \citep{arnott_artificially_2021}. We
caution that determination of the shape of the nutrient depletion
volume in the presence of turbulence is too complex to be addressed
here \citep{karp-boss_nutrient_1996}.

\medskip{}

We consider a total volume of 1000 cm$^{3}$ for microphytoplankton
and 10 cm$^{3}$ for nanophytoplankton (volumes are adapted to balance
realistic concentrations and computation time) with periodic boundary
conditions. Individuals are uniformly distributed in the cube at the
beginning of the simulation. We run an idealized simulation with 3
species with an even abundance distribution of about $10^{4}$ cells~L$^{-1}$
for microphytoplankton \citep{picoche_strong_2020} and $10^{6}$
cells~L$^{-1}$ for nanophytoplankton individuals \citep{edwards_mixotrophy_2019}.
We then model a more realistic community with 10 species having a
skewed abundance distribution (between 55,000 and 400 cells~L$^{-1}$
for microphytoplankton, according to observations of field abundance
distributions in \citealp{picoche_strong_2020}, and multiplied by
$10^{2}$ for nanophytoplankton). All simulations are run for 1000
time steps of duration $\tau$ (corresponding to approximately 6h40 of phytoplankton time---note that runtimes can be much longer). The computation of $g$ and $K$ for simulated distributions is explained in Section~\ref{section:si_computation} of the SI. The code for all simulations and analyses can be found at \url{https://github.com/CoraliePicoche/brownian_bug_3D/}.

\section*{Results}

\begin{figure}[H]
\begin{centering}
\includegraphics[width=0.99\textwidth]{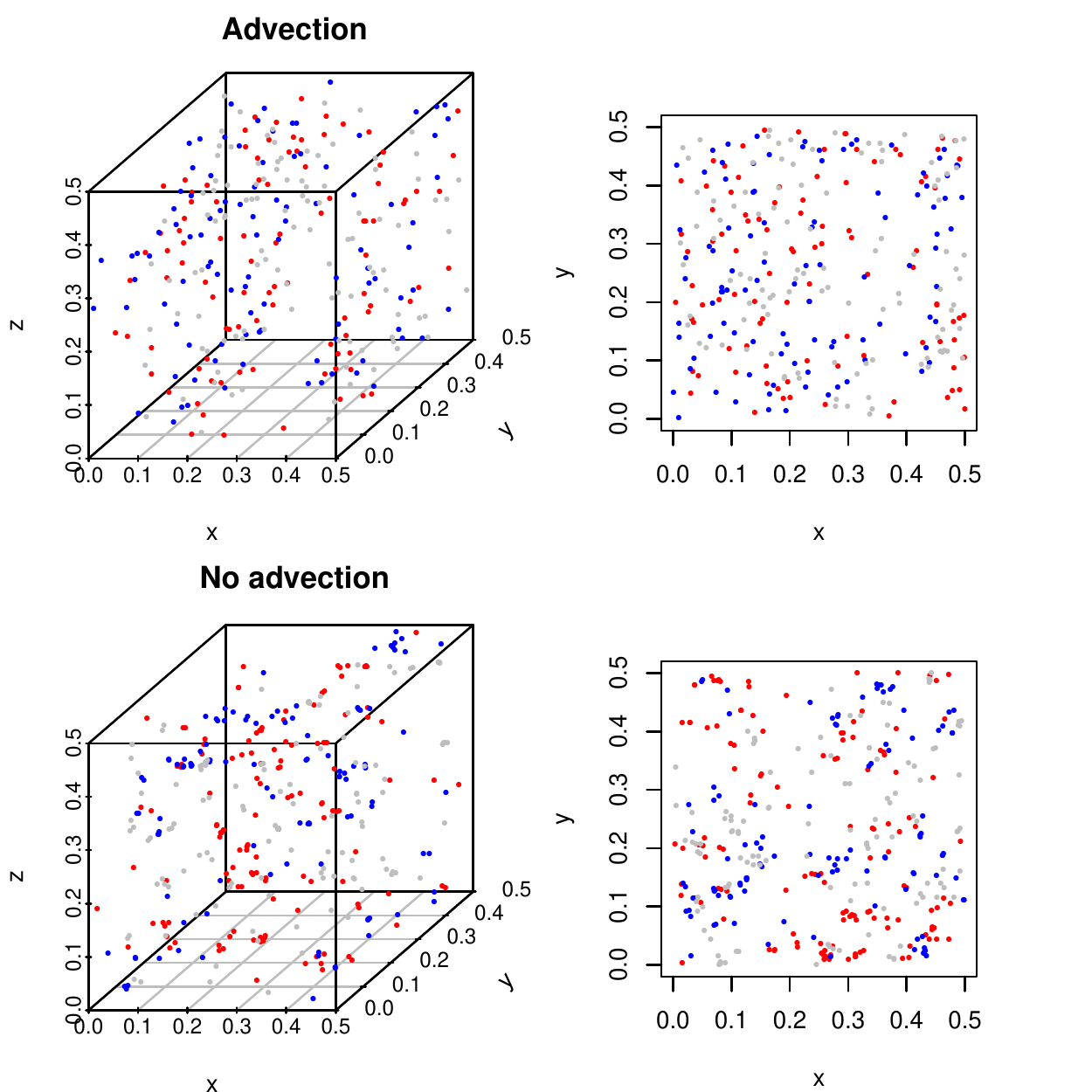} 
\par\end{centering}
\caption{Spatial distributions of a 3-species community of nanophytoplankton
with and without advection with density $C=10^{3}$ cells~cm$^{-3}$
after 1000 time steps. Each color corresponds to a different species.
On the left-hand side, only a zoom on a $0.5\times0.5\times0.5$ cm$^{3}$
cube is shown, and its projection on the x-y plane is shown on the
right-hand side. \label{fig:Spatial-distributions}}
\end{figure}

We show an example of nanophytoplankton spatial distributions with
and without advection at the end of a simulation in Fig. \ref{fig:Spatial-distributions}:
clustering is not visible to the naked eye, even when zooming in on
the observation volume, in the presence of advection, but removing
turbulence helps visualising small aggregates of conspecifics. Microphytoplankton distributions are not straightforward to interpret as no clusters can be detected visually (although they may actually be present), whether advection is included or not (Section~\ref{section:si_spatial_distrib} of the SI). Statistics are therefore needed to go further in detecting patterns of aggregation.

\begin{figure}[H]
\begin{centering}
\includegraphics[width=0.99\textwidth]{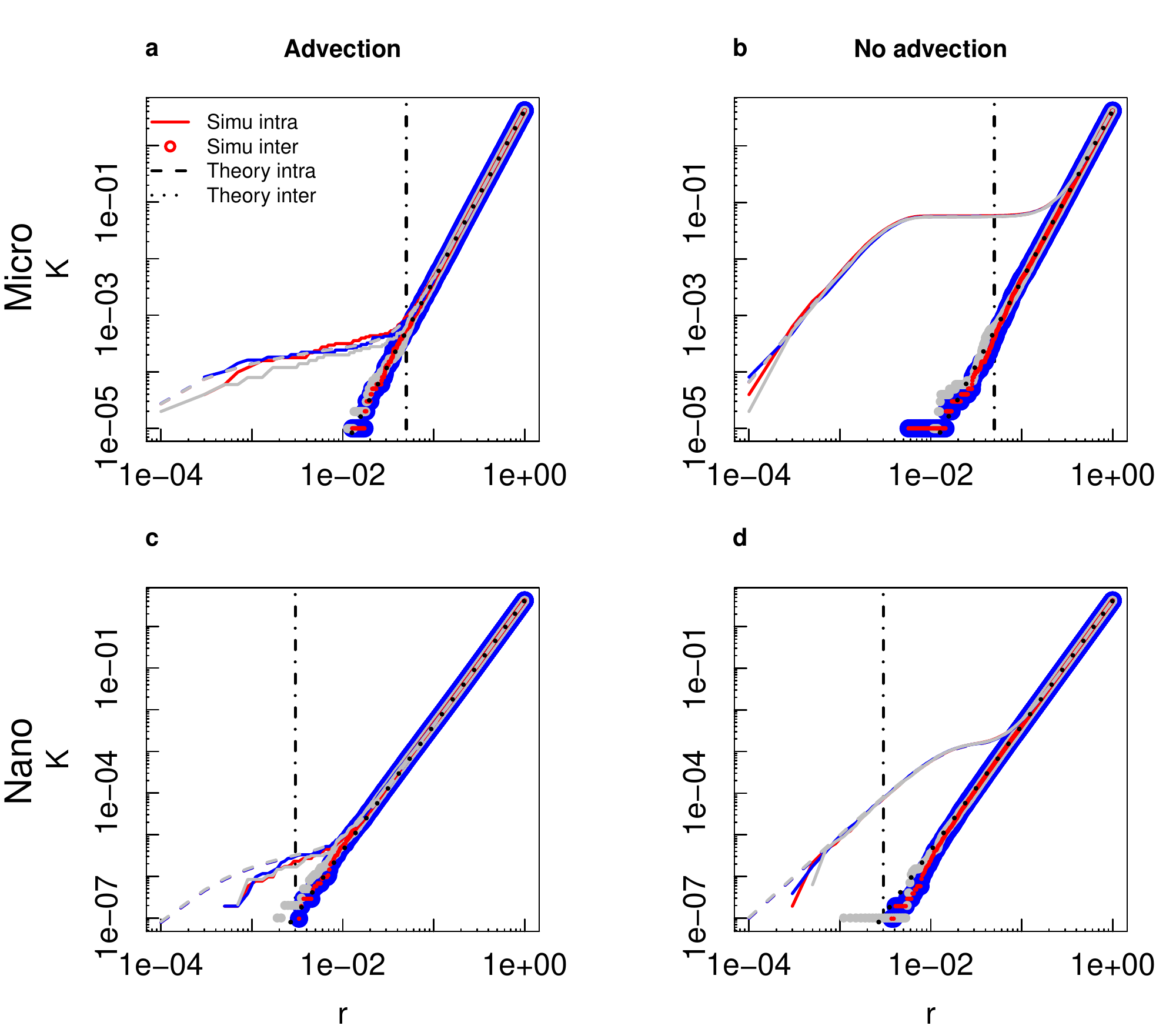} 
\par\end{centering}
\caption{Comparison of theoretical and simulated Ripley's $K$-functions as
a function of distance (in cm) for microphytoplankton (a-b) and nanophytoplankton
(c-d) in a 3-species community with even abundance distributions after
1000 timesteps, with (a, c) and without (b, d) advection. Each color
represents a different species. Intraspecific $K$-functions are shown
with dashed (theoretical values) and solid (simulated values) lines.
Interspecific $K$-functions are shown with dotted lines (theoretical
values) and circles (simulated values). The black dash-dotted line
corresponds to the threshold considered as the maximum distance for
nutrient-based competition.\label{fig:Ripley's-K-function}}
\end{figure}

Ripley's $K$-functions extracted from numerical simulations match
theoretical formula (Fig. \ref{fig:Ripley's-K-function}) for both
types of organisms, which also indicates that dominance indices extracted
from the simulations match theoretical expectations.

\begin{figure}[H]
\begin{centering}
\includegraphics[width=0.99\textwidth]{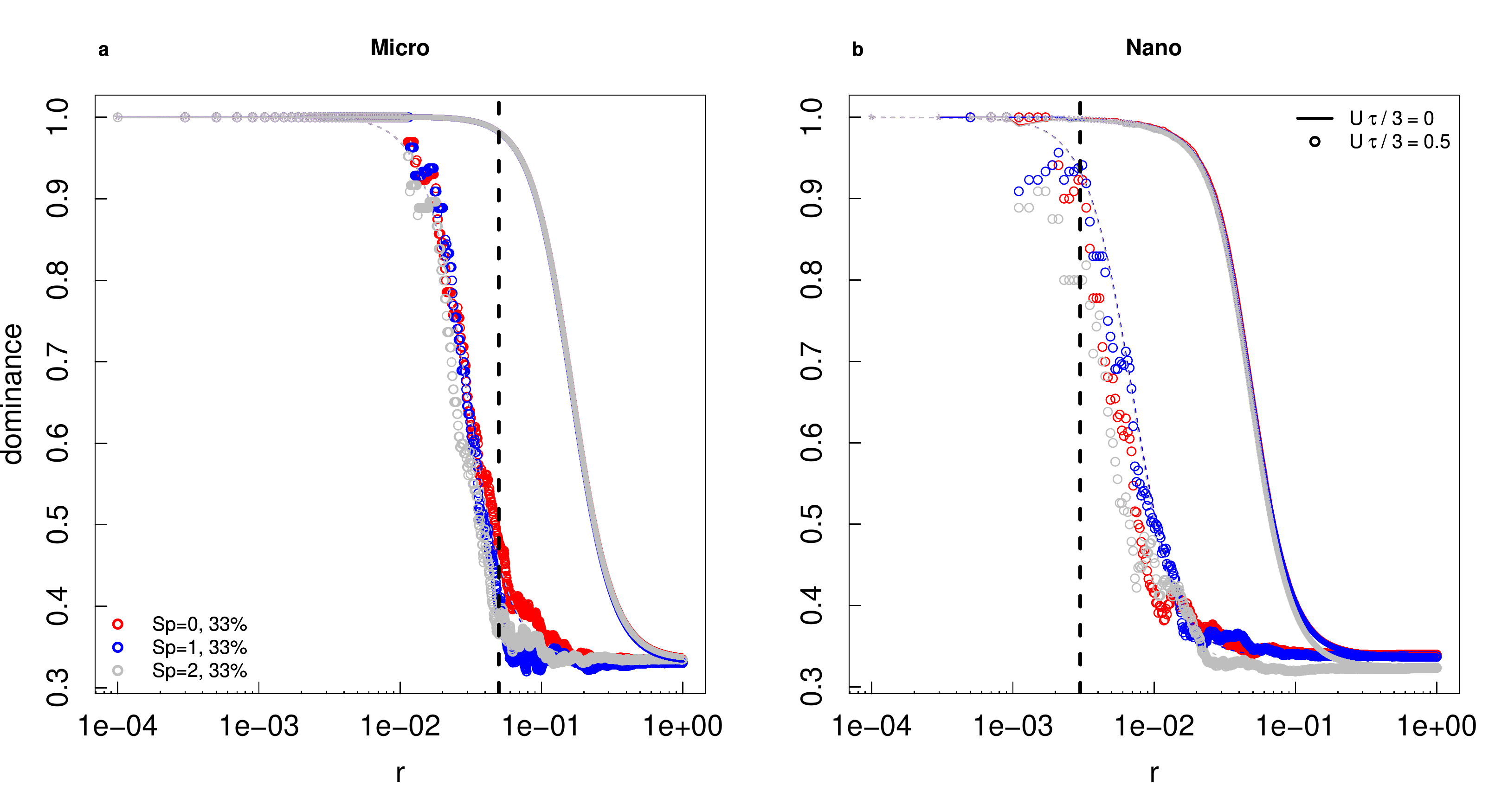} 
\par\end{centering}
\caption{Dominance indices as a function of distance (in cm) for microphytoplankton
(a) and nanophytoplankton (b) in a 3-species community with even abundance
distributions (final proportions in the community are indicated in
the figure) after 1000 timesteps, with (circles) and without (lines)
advection. Each color represents a different species. The grey dashed curve represents the analytical solution. The black dashed line corresponds to the threshold considered as the maximum distance for nutrient-based competition.\label{fig:Dominance-3sp}}
\end{figure}

\begin{figure}[H]
\begin{centering}
\includegraphics[width=0.99\textwidth]{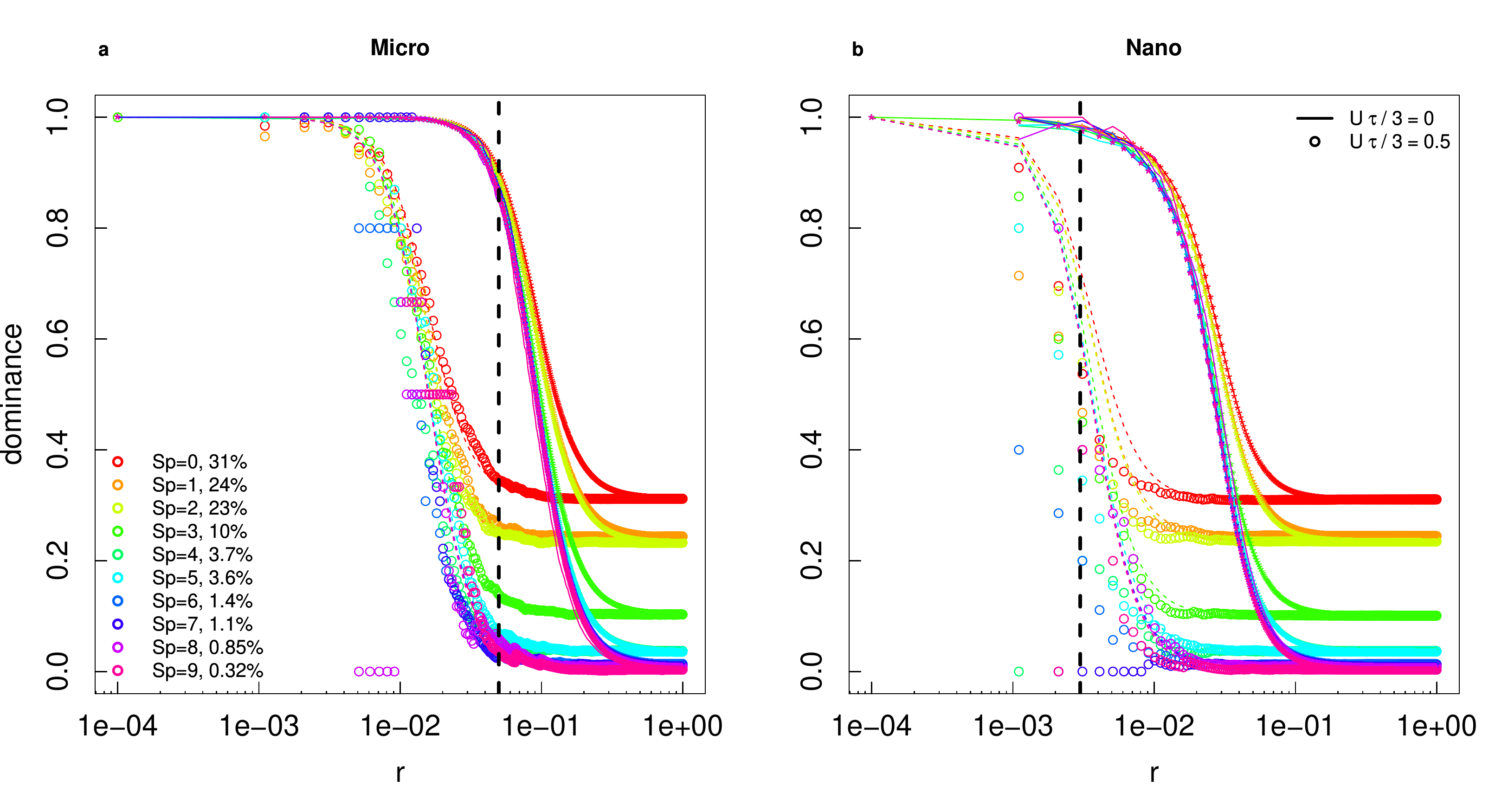} 
\par\end{centering}
\caption{Dominance indices as a function of distance (in cm) for microphytoplankton
(a) and nanophytoplankton (b) in a 10-species community with a skewed
abundance distribution (final proportions in the community are indicated
in the figure) after 1000 timesteps, with (circles) and without (lines)
advection. Each color represents a different species. The coloured dashed curves (advection) and small stars (no advection) represent the analytical solution. The black dashed line corresponds to the threshold considered as the maximum distance for nutrient-based competition.\label{fig:Dominance-10sp}}
\end{figure}

Dominance indices all follow a similar pattern (Fig. \ref{fig:Dominance-3sp}
and \ref{fig:Dominance-10sp}). The dominance index is close to 1
for small distances: there is always a scale at which an organism
is surrounded almost only by conspecifics. The index then decreases
sharply to converge at large distances (close to 1 cm) to the proportion
of the focus species in the whole community, as it would for a uniform
spatial distribution. Patterns differ at intermediate ranges of distances
between organisms.

In the presence of advection, the dominance index starts decreasing
for a distance approximately 10 times smaller than when advection is absent, which indicates that organisms are closer to heterospecifics when their environment is turbulent. A quasi-uniform distribution is also reached for smaller distances with advection than without. Microphytoplankton species start mixing for distances larger than for nanophytoplankton species irrespective of the hydrodynamic regime surrounding them. 

In a 3-species community with the same initial abundances, in the
presence of advection, microphytoplankton dominance indices are between
0.37 and 0.47 at the distance threshold for potential interactions,
while they are between 0.80 and 0.94 for nanophytoplankton species.
In the absence of turbulence, dominance indices are all above 0.98
when the distance threshold is reached (Fig. \ref{fig:Dominance-3sp}).
Microphytoplankton organisms are therefore as likely to share their
depletion volume with conspecifics as they are with heterospecifics,
but only when turbulent advection is accounted for, whereas nanophytoplankton
organisms always have almost only conspecifics around them.

More mixing in microphytoplankton than nanophytoplankton, and more
mixing with advection, also holds when considering a 10 species-community
with a skewed abundance distribution (Fig. \ref{fig:Dominance-10sp}),
but dominance indices are overall lower in communities with more species
and with less even abundances. In the presence of advection, microphytoplankton
dominance indices at the distance threshold are between 0.34 (for
the most abundant species) and 0.033 (for one of the least abundant
species), while they are between 0.90 and 0.85 when advection is not
taken into account. Nanophytoplankton species, too, are more mixed
than in the 3 species-community: dominance indices vary between 0.54
and 0.2 when the depletion threshold is reached (with an exception
of 0 for one particular species which had no conspecific for distances
below $10^{-2}$ cm) when organisms are displaced by turbulence, while
the same quantity is between 1 and 0.97 when they are only subject
to diffusion.

\begin{figure}[H]
\begin{centering}
\includegraphics[width=0.8\textwidth]{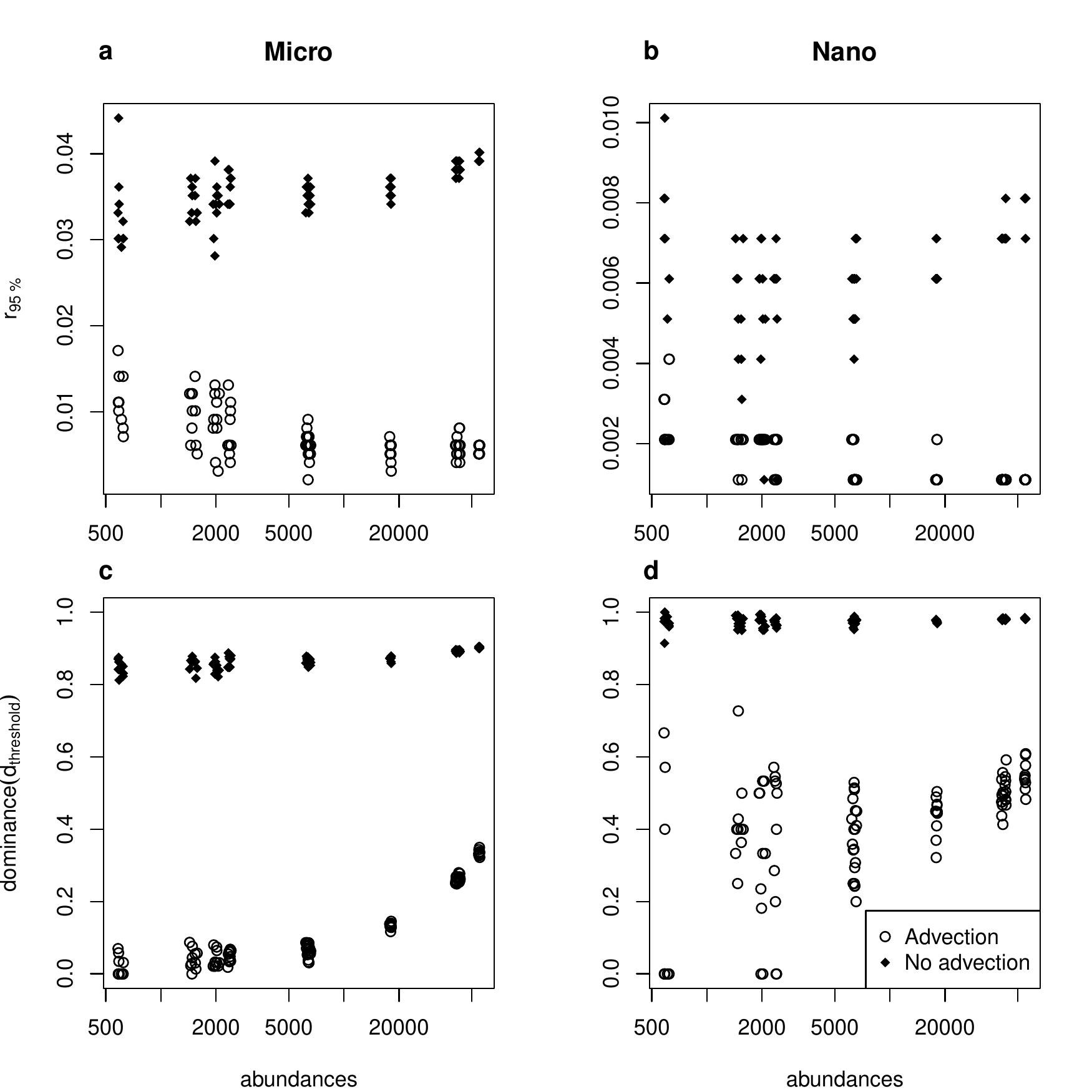} 
\par\end{centering}
\caption{Minimum distances (in cm) between points for dominance to drop below
95\% (a and b) and dominance at a distance corresponding to the threshold
for competition (c and d) as a function of abundances (note the logarithmic
scale on the x-axis) for microphytoplankton and nanophytoplankton.
We consider cases with and without advection in a 10-species community
with a skewed abundance distribution. These have been obtained combining 10 sets of simulations. \label{fig:Carac_10sp}}
\end{figure}

Differences in spatial distributions are not only due to organism
sizes, which determine their demographic and hydrodynamic properties,
but also to their abundances (here set through initial values). In
the presence of turbulence, the threshold distance at which dominance
falls below 95\% is smaller for more abundant species (Fig. \ref{fig:Carac_10sp}
a-b). Abundant species tend to be present nearly everywhere when they
are mixed in the environment. Therefore, they are also more likely
to be close to a heterospecific, but still have more conspecifics
close to them than the less abundant species ($\mathcal{D}\left(d_{\text{threshold}}\right)$
increases with abundance, Fig. \ref{fig:Carac_10sp} c-d). However,
this increase is less clear for nanophytoplankton than for microphytoplankton
(Fig. \ref{fig:Carac_10sp} c-d). When turbulence is absent, the relationships
with abundance are unclear, possibly affected by sampling effects,
and we refrain from interpreting them.

\section*{Discussion}

We designed a stochastic, three-dimensional, individual-based model
of the spatial distribution of multiple species in a viscous and turbulent
flow. We conducted both mathematical analyses and numerical simulations
to quantify spatial correlations in the distribution of organisms.
We focused on the pair correlation function and Ripley's $K$-function,
for which numerical and theoretical analyses showed a good agreement,
and extracted a more ecologically-oriented metric from them, i.e.,
the dominance index. This statistic is the \textit{local} average
ratio of conspecifics, i.e., the number of organisms of the focal
species in the neighbourhood of an individual of the same species,
divided by the total number of organisms in that neighbourhood. Intraspecific
clustering corresponds to a dominance index close to 1, which decreases
when interspecific mixing increases. The choice of this index was
motivated by two reasons: (1) it is at its core a proportion of a
focal species in a certain volume, i.e. a scale-dependent, localized metric bounded between 0 and 1 as opposed to other statistics whose values are less directly interpreted, and (2) it is easy to relate to coexistence theory as it describes the environment of an organism
in terms of heterospecifics and conspecifics, which can, under certain
assumptions that we discuss below, be related to interspecific and intraspecific interactions. Comparing the distributions of organisms of different sizes, we showed that the presence of turbulence always increased mixing (results are robust to slight modifications in the
computation of advection velocity $U$, shown in Section~\ref{section:si_advection_param} of the
SI). The species composition around an organism depended on its size,
which mechanically determines its hydrodynamic properties (diffusivity),
and is linked with its ecological characteristics (growth rate and
density). Microphytoplankters (20 to 200 \textmu m), larger cells
with lower diffusivity, growth rate and abundance, were on average
further away from other cells, due to their lower concentrations (Fig.~\ref{fig:Distance_abundance} of the SI), than nanophytoplankters (2 to 20 \textmu m). They were surrounded by more heterospecifics than conspecifics within
a volume of potential interactions, whose radius is defined as the
maximum distance for which nutrient depletion volumes of two different
individuals may overlap. If we consider that interactions between
species (not modelled directly here because of timescale issues, see
below) could occur with equal probability at all distances within
the volume of potential interactions, we would conclude that microphytoplankters are more likely to interact with individuals from other species than with individuals of their own species. This affirmation is, however, conditional upon interactions at 10 cell diameters from an individual being equally likely than at 1 diameter from an individual. If we
keep in mind that interactions are more likely or stronger at very
short distances, or that the maximal radius of interaction could be shorter than our estimation and advection velocity $U$ lower (SI Section~\ref{section:si_advection_param}), microphytoplankters may still experience more frequent effects of conspecifics than heterospecifics.

To see this, let us first focus on the smallest distances between
organisms. The nearest neighbour of an organism was always an organism
of the same species, and the minimum distance between conspecifics
was always lower than expected for a uniform distribution (Section~\ref{section:si_min_distances} of the SI). The dominance index remained close to 1 for distances
below $10^{-2}$ cm or $10^{-3}$ cm for microphytoplankton and nanophytoplankton
respectively. There was therefore always \emph{some} intraspecific
aggregation, i.e., conspecifics were always closer than heterospecifics
at the smallest distances. This is due to the prevalence of demographic
processes at individual scales, because an individual acts as a source
point for other organisms of the same species, and hydrodynamic processes
do not separate conspecifics fast enough to prevent aggregation. This remains true if we add an initial separation distance between mother and daughter cells upon birth (additional simulations, see code repository). If we consider that interaction strengths are a smoothly decaying function
of distance, a common assumption in spatial coexistence models \citep[e.g., ][]{bolker_spatial_1999,law_population_2003},
this implies that population-level intraspecific interactions could
be stronger than interspecific interactions due to intraspecific micro-scale
aggregation. However, the mechanisms of competition at this scale
are poorly known, likely relying on multiple types of resources with
different distributions in the environment, effects on the cell, uptakes,
etc. Rather than weighting much more heavily the potential interactions
with the closest neighbour(s) through an interaction kernel, we therefore
chose conservatively to define a maximum distance for two organisms
to possibly affect the concentrations of elements in the environment
of each other, assuming perfect absorption on the cell surface. We consider that, at all distances below this threshold, interactions could happen between organisms. We continue the discussion with that simplification in mind, and explicitly mention when it is relaxed.

Dominance indices began to decrease at distances above $10^{-3}$
cm, still below the maximum distance for interactions. At this distance
and above, the balance between heterospecifics and conspecifics was
much more sensitive to different phytoplankters' demographic and hydrodynamic
traits. The species composition of an organism's neighbourhood depended
on its size: nanophytoplankton organisms mainly shared their volume
of potential interactions with conspecifics (the dominance index remained
close to 1, even near the distance threshold, i.e., the maximum distance
for the overlap of nutrient depletion volumes) while microphytoplankton
organisms could affect both conspecifics and heterospecifics (the
dominance index was often below 0.5 at the distance threshold, i.e.,
an individual's depletion zone probably overlapped with more heterospecifics'
than conspecifics'). Microphytoplankters were therefore more likely
to share their depletion volume with heterospecifics than nanophytoplankters.
The rate of production of new microphytoplankton conspecifics was
not sufficient to compensate for the mixing induced by turbulence
and diffusivity, even though the diffusivity of microphytoplankters
was smaller than that of nanophytoplankters. There may therefore be
different mechanisms at play at the community level for microphytoplankton
and nanophytoplankton to maintain coexistence. For nanophytoplankton,
the spatial structure likely leads to more interactions between conspecifics
than between heterospecifics. The spatial distribution of microphytoplankton
species, on the contrary, encourages more interactions between heterospecifics.
If we consider that local interaction strengths are equal within the
volume of potential interactions, scaling to the population level,
we would likely observe stronger intra- over interspecific interactions
for nanophytoplankton (a key factor in coexistence theory, \citealp{barabas_self-regulation_2017})
but not necessarily so for microphytoplankton. Using a timescale separation
argument, we show in Section~\ref{section:si_pop_interactions} in the SI how stronger interactions
at population level than individual level may arise in a Lotka-Volterra
model whose spatial structure is summed up by the dominance indices
evidenced here. Stronger intra- than interspecific competition may
arise at population level even when assuming that all local interaction
strengths between individuals are equal, regardless of the identity
of competitors.

All of the above discussion is based on a microphytoplankter's neighbourhood
in its nutrient depletion volume. To simplify the computation, we
used maximum volumes of potential interactions, corresponding to a
diffusive-only flow of nutrient particles. But when fluid turbulence
increases, nutrient uptake increases, and the size of the depletion
zone decreases \citep{karp-boss_nutrient_1996}. The proportion of
change in the depletion volume increases with the size of organisms:
a 10 \textmu m-diameter organism might not experience any change,
while the uptake of a 100 \textmu m-diameter organism would increase
by at least 50\% \citep{karp-boss_nutrient_1996}. Therefore the volume
of potential interactions shrinks in the presence of turbulence for
microphytoplankton, but not necessarily for nanophytoplankton. An additional reason why microphytoplankers might still
be surrounded by conspecifics at ecologically meaningful distances
and interacting more frequently with them is imperfect absorption of nutrients: if nutrient concentration at the cell surface is not zero but $C_0$, then the radius of interaction is $10 a_i(1 - C_0/C_{\infty})$. 

Up to now, we have only focused on the dominance index, a localized
proportion of conspecifics. However, interactions also depend on the
absolute densities of individuals. Mechanically, when density decreases,
the distances between neighbours increase, which explains that the
distances between the low-abundance microphytoplankters tended to
be greater than distances between the more abundant nanophytoplankters
(Section~\ref{section:si_min_distances} of the SI). Explicit mathematical models using pair densities
to express interaction rates \citep[e.g.][]{law_population_2003,plank_spatial_2015}
may be able to incorporate those effects; however, as we highlight
below, the timescales and spatial correlations that are seen in such
models may not necessarily represent faithfully phytoplankton community
dynamics. 

Contrary to other similar models \citep[e.g.,][]{birch_master_2006,bouderbala_3d_2018},
we did not consider explicit effects of local density on survival
and fertility rates. Outside of simply maintaining analytical tractability,
we had another, more biological reason to do so: we cannot be sure
that these local density-dependencies make sense in our phytoplankton
context. To understand why, consider that even if a species abundance
is locally tripled, competition might not directly ensue at the time
scales covered by our model ($\approx7$ h), if nutrient depletion
has not had time to set in yet. Even if we considered longer time
frames, we would need lagged local density-dependencies, which are
to our knowledge not leading to tractable spatial branching or dynamic
point processes. We could, of course, directly model nutrients, perhaps
as resource ``points'' with a dynamics of their own \citep{murrell_local_2005,north_interactions_2007},
which in turn change the reproduction or death rate of individuals.
If the resource points risk being depleted, this entails a negative
spatial correlation between organisms and their resources \citep{murrell_local_2005,barraquand2012evolutionarily}.
And that is where such models might be inadequate. The phycosphere,
a micro-environment at the periphery of a phytoplankton organism where
communities of bacteria interact \citep{seymour_zooming_2017}, can
also impact phytoplankton fitness, both positively (cross-feeding)
and negatively (algicidal activities of bacteria). This can sometimes
lead to an accumulation of key resources close to the phytoplankter.
This will lead to positive spatial correlations between consumers
and their resources, and we currently do not have theoretical models
to represent this process (short of modelling precisely the spatial
distribution of these bacteria).

Our model should be viewed as a first model of spatial distributions
of multiple phytoplankton species in a realistic, three-dimensional
environment at the microscale, describing only basic hydrodynamic
and demographic processes. Using this model, we were able to predict
whether phytoplankters could be in contact with individuals of their
own or other species, and form reasonable conjectures regarding potential intra vs interspecific interactions between species, emerging at the population level through spatial distributions \citep{detto_stabilization_2016}.
It is worthwhile to keep in mind that there are many remaining features
of phytoplankton physiology and life histories which we do not address
here, but which may affect spatial distributions. Many phytoplankters
are able to move actively in three dimensions, which can favour cluster
formation \citep{breier_emergence_2018}. Even those who are believed
to move passively actually often move along the vertical dimension
by regulating their buoyancy \citep{reynolds2006ecology}, and can
at times aggregate to form pairs \citep{font-munoz_collective_2019}.
Finally, a part of spatial structure is explained by the partially
colonial nature of microphytoplankton \citep{kiorboe_coagulation_1990}.
This clearly calls for viewing our model as a null model to which
more complex mechanistic models and their spatial outputs can be compared.

\subsubsection*{Acknowledgements}

FB and CP were supported by the grant ANR-20-CE45-0004. CP was supported
by a PhD grant from the French Ministry of Research. We are thankful for constructive reviewer feedback. 

\section*{Appendices}

\global\long\def\thesection{A\arabic{section}}%
 \setcounter{section}{0} 
\global\long\def\thefigure{A\arabic{figure}}%
 \setcounter{figure}{0} 
\global\long\def\thetable{A\arabic{table}}%

\subsection*{Derivation of the spatial characteristics of the Brownian Bug Model}

We show here how to compute the monospecific pair correlation function
and Ripley's $K$-function of the Brownian Bug Model (see \citealp{young_reproductive_2001}
and \citealp{picoche_rescience_2022} for a detailed derivation from
the master equation). As these formula only apply to intraspecies
pairs, we ignore species' index in the following for the sake of clarity.
Similar formula for well-known spatial point processes are given in
the Supplementary Information, for readers who want to understand
better the properties of these spatial statistics.

\subsubsection*{Proof of Eq. \ref{eq:pcf_intra_adv} and Eq. \ref{eq:pcf_intra_noadv}}

In three dimensions, when the birth rate $\lambda$ is the same as
the mortality rate $\mu$, the pair density $G(r)$ is a solution
of 
\begin{equation}
\frac{\partial G}{\partial t}=\frac{2D}{r^{2}}\frac{\partial}{\partial r}\left(r^{2}\frac{\partial G}{\partial r}\right)+\frac{\gamma}{r^{2}}\frac{\partial}{\partial r}\left(r^{4}\frac{\partial G}{\partial r}\right)+2\lambda C\delta(\boldsymbol{\xi}).\label{eq:Young3D-1}
\end{equation}

\paragraph*{Steady-state solution}

We first compute the steady-state solution,\textit{ i.e.} 
\begin{align}
\,0 & =\frac{2D}{r^{2}}\frac{\partial}{\partial r}\left(r^{2}\frac{\partial G}{\partial r}\right)+\frac{\gamma}{r^{2}}\frac{\partial}{\partial r}\left(r^{4}\frac{\partial G}{\partial r}\right)+2\lambda C\delta(\boldsymbol{\xi})\nonumber \\
0 & =4\pi r^{2}\left(\frac{2D}{r^{2}}\frac{\partial}{\partial r}\left(r^{2}\frac{\partial G}{\partial r}\right)+\frac{\gamma}{r^{2}}\frac{\partial}{\partial r}\left(r^{4}\frac{\partial G}{\partial r}\right)+2\lambda C\delta(\boldsymbol{\boldsymbol{\xi}})\right)\nonumber \\
0 & =4\pi\left(2D\frac{\partial}{\partial r}\left(r^{2}\frac{\partial G}{\partial r}\right)+\gamma\frac{\partial}{\partial r}\left(r^{4}\frac{\partial G}{\partial r}\right)\right)+4\pi r^{2}2\lambda C\delta(\boldsymbol{\boldsymbol{\xi}}).\label{eq:steady_state}
\end{align}
We can then integrate Eq. \ref{eq:Young3D-1} over a small sphere
centered on an individual, with radius $\rho$. Let us first note
that in an isotropic environment the 3D-Dirac function relates to the radial one as
\begin{align}
 \delta(\boldsymbol{\xi}) =  \frac{1}{4 \pi r^2}\delta(r) \label{eq:delta_integration}
\end{align}
with $4\pi r^2$ the surface of the sphere of radius $r$.
Using Eq. \ref{eq:steady_state} and \ref{eq:delta_integration},
\begin{align}
0 & =4\pi\left(2Dr^{2}\frac{\partial G}{\partial r}+\gamma r^{4}\frac{\partial G}{\partial r}\right)+2\lambda C\nonumber \\
\Leftrightarrow\frac{\partial G}{\partial r} & =-\frac{1}{4\pi}\frac{2\lambda C}{2Dr^{2}+\gamma r^{4}}.\label{eq:deriv_G_r}
\end{align}
We can integrate Eq. \ref{eq:deriv_G_r} between $\rho$ and $\infty$.
As $G(\infty)=C^{2},$ 
\begin{align}
C^{2}-G(\rho) & =-\frac{\lambda C}{2\pi}{\displaystyle \int_{\rho}^{\infty}}\frac{1}{2Dr^{2}+\gamma r^{4}}dr.\label{eq:deriv_G_r_int1}
\end{align}
We first compute the primitive $A=\int\frac{1}{2Dr^{2}+\gamma r^{4}}dr$.
\begin{align}
A & =\int\frac{1}{r^{2}\left(2D+\gamma r^{2}\right)}dr\\
 & =\int\frac{1}{2Dr^{2}}-\frac{\gamma}{2D\left(2D+\gamma r^{2}\right)}dr\\
 & =-\frac{1}{2Dr}-\frac{\gamma}{2D}\int\frac{1}{2D\left(1+\left(\sqrt{\frac{\gamma}{2D}}r\right)^{2}\right)}dr.
\end{align}
With a change of variable $u=\sqrt{\frac{\gamma}{2D}}r$, using $\int\frac{1}{1+u^{2}}du =\arctan(u)$,
we have 
\begin{equation}
A=-\frac{1}{2Dr}-\frac{\sqrt{\gamma}\arctan\left(\frac{\sqrt{\gamma}r}{\sqrt{2D}}\right)}{2\sqrt{2}D\sqrt{D}}+K
\end{equation}
where $K$ is a constant. We can now compute $B=[A]_{\rho}^{\infty}.$
\begin{equation}
B=-\frac{\sqrt{\gamma}\pi}{4\sqrt{2}D\sqrt{D}}+\frac{1}{2D\rho}+\frac{\sqrt{\gamma}\arctan\left(\frac{\sqrt{\gamma}\rho}{\sqrt{2D}}\right)}{2\sqrt{2}D\sqrt{D}}.
\end{equation}
This leads to 
\begin{align}
G(\rho)= & C^{2}+\frac{\lambda C}{2\pi}B\\
= & C^{2}+\frac{\lambda C}{2\pi}\left[\frac{1}{2D\rho}+\frac{\sqrt{\gamma}\arctan\left(\frac{\sqrt{\gamma}\rho}{\sqrt{2D}}\right)}{2\sqrt{2}D\sqrt{D}}-\frac{\sqrt{\gamma}\pi}{4\sqrt{2}D\sqrt{D}}\right].
\end{align}
Finally, the pair correlation function $g=G/C^{2}$ is defined as
\begin{equation}
g(\rho)=\frac{\lambda}{4\pi CD}\left(\frac{\sqrt{\gamma}\arctan\left(\frac{\sqrt{\gamma}\rho}{\sqrt{2D}}\right)}{\sqrt{2D}}+\frac{1}{\rho}-\frac{\pi\sqrt{\gamma}}{2\sqrt{2D}}\right)+1.\label{eq:pcf_adv_bbm}
\end{equation}

\paragraph{Time-dependent solution}

In the absence of advection by turbulent diffusion ($U=0,\,\gamma=0$),
convergence to the steady-state solution can be very slow (more than
a week, see Section~\ref{section:si_convergence} in the SI). In order to keep a realistic timeframe,
we need to compute a time-dependent solution. We can get back to Eq.
\ref{eq:Young3D-1} with $\gamma=0$, which yields 
\begin{equation}
\frac{\partial G}{\partial t}=\frac{2D}{r^{2}}\frac{\partial}{\partial r}\left(r^{2}\frac{\partial G}{\partial r}\right)+2\lambda C\delta(\boldsymbol{\xi}).\label{eq:G_no_adv}
\end{equation}
Assuming an isotropic environment, this means 
\begin{equation}
\frac{\partial G}{\partial t}-2D\Delta G=2\lambda C\delta(\boldsymbol{\xi})
\end{equation}
where $\Delta=\nabla^{2}$ is the Laplacian operator. We therefore
have 
\begin{equation}
\mathcal{L}G(\boldsymbol{\xi},t)=2\lambda C\delta(\boldsymbol{\xi})
\end{equation}
where $\mathcal{L}$ is the linear differential operator $\partial_{t}-2D\Delta$. Therefore $G(y)=\int H(y,s)2\lambda C\delta(s)ds$ where $H(y,s)=H(y-s)$ is the Green kernel (heat kernel). We can therefore write 
\begin{equation}
\begin{array}{ccc}
G(\boldsymbol{\xi},t) & = & 2\lambda C\int_{\mathbb{R}^{3}}\int_{0}^{t}H(\boldsymbol{\xi}-\boldsymbol{\xi}',t')\delta(\boldsymbol{\xi}')d\boldsymbol{\xi}'dt'\\
\Leftrightarrow G(\boldsymbol{\xi},t) & = & 2\lambda C\int_{0}^{t}H(\boldsymbol{\boldsymbol{\xi}},t')dt'.
\end{array}
\end{equation}

A solution for the Green's function using $\mathcal{L}=\partial_{t}-2D\Delta$
in three dimensions is $H(r,t)=\left(\frac{1}{8\pi Dt}\right)^{3/2}\exp(\frac{-r^{2}}{8Dt})$.
$G(r,t)$ can then be computed as 
\begin{equation}
G(r,t)=2\lambda C\left(\frac{-\erf\left(\frac{r}{\sqrt{8tD}}\right)}{8\pi Dr}+K\right)\label{eq:G_r_t}
\end{equation}
where $\erf$ is the error function. Using $G(r,0)=C^{2}$ and $\lim_{x\rightarrow+\infty}\erf(x)=1$
in Eq. \ref{eq:G_r_t}, 
\begin{equation}
\begin{array}{ccc}
C^{2} & = & 2\lambda C\left(\frac{-1}{8\pi Dr}+K\right)\\
\Leftrightarrow\frac{C}{2\lambda}+ & \frac{1}{8\pi Dr}= & K.
\end{array}
\end{equation}
We can finally compute $G(r,t)$: 
\begin{equation}
\begin{array}{ccc}
G(r,t) & = & 2\lambda C\left(-\frac{\erf\left(\frac{r}{\sqrt{8tD}}\right)}{8\pi Dr}+\frac{C}{2\lambda}+\frac{1}{8D\pi r}\right)\\
 & = & \frac{\lambda C}{4\pi Dr}\left\{ 1-\erf\left(\frac{r}{\sqrt{8Dt}}\right)\right\} +C^{2}\\
\Leftrightarrow g(r,t) & = & \frac{\lambda}{4D\pi rC}\left\{ 1-\erf\left(\frac{r}{\sqrt{8Dt}}\right)\right\} +1.
\end{array}\label{eq:pcf_noadv_bbm}
\end{equation}

\subsubsection*{Proof of Eq. \ref{eq:K_intra_adv} and Eq. \ref{eq:K_intra_no_adv}}

We can integrate the pcf formula to compute Ripley's $K$-function,
as $g(r)=\frac{K'(r)}{4\pi r^{2}}$.

\paragraph{Steady-state solution}

From Eq. \ref{eq:pcf_adv_bbm}, 
\begin{equation}
\begin{array}{ccc}
K(\rho) & = & 4\pi\int_{0}^{\rho}r^{2}+\frac{\lambda}{2\pi C}\left[\frac{r}{2D}+\frac{\sqrt{\gamma}r^{2}\arctan\left(\frac{\sqrt{\gamma}r}{\sqrt{2D}}\right)}{2\sqrt{2}D\sqrt{D}}-\frac{\sqrt{\gamma}\pi r^{2}}{4\sqrt{2}D\sqrt{D}}\right]dr.\end{array}\label{eq:start-K}
\end{equation}
We define $A=\int_{0}^{\rho}r^{2}dr$ , $B=\int_{0}^{\rho}\frac{r}{2D}dr$,
$C=\int_{0}^{\rho}r^{2}\arctan\left(\frac{\sqrt{\gamma}r}{\sqrt{2D}}\right)dr$
and $E=\int_{0}^{\rho}\frac{\sqrt{\gamma}\pi r^{2}}{4\sqrt{2}D\sqrt{D}}dr$.
\begin{equation}
\begin{array}{cc}
A= & \frac{1}{3}\rho^{3}.\\
B= & \frac{\rho^{2}}{4D}.\\
E= & \frac{\sqrt{\gamma}\pi\rho^{3}}{12\sqrt{2}D\sqrt{D}}.
\end{array}
\end{equation}
We can also compute $C=\int_{0}^{\rho}r^{2}\arctan\left(\frac{\sqrt{\gamma}r}{\sqrt{2D}}\right)dr$.
We first change variable, with $u=\frac{r}{\sqrt{2D}}$, $dr=\sqrt{2D}du$,
and obtain 
\begin{equation}
\begin{array}{ccc}
C & = & (2D)^{3/2}\int_{0}^{\rho/\sqrt{2D}}u^{2}\arctan(\sqrt{\gamma}u)du.\end{array}
\end{equation}
We can integrate by parts, with $f=\arctan(\sqrt{\gamma}u)$ and $g'=u^{2}$,
which leads to 
\begin{equation}
\begin{array}{ccc}
C & = & (2D)^{3/2}\left(\frac{\rho^{3}}{3(2D)^{3/2}}\arctan(\sqrt{\frac{\gamma}{2D}}\rho)-\frac{\sqrt{\gamma}}{3}\int_{0}^{\rho/\sqrt{2D}}\frac{u^{3}}{(\gamma u^{2}+1)}du\right).\end{array}
\end{equation}
We then substitute $v=\gamma u^{2}+1$, $du=\frac{1}{2\gamma u}dv$,
and have 
\begin{equation}
\begin{array}{ccc}
\int_{0}^{\rho/\sqrt{2D}}\frac{u^{3}}{(\gamma u^{2}+1)}du & = & \frac{1}{2\gamma^{2}}\int_{1}^{\gamma\rho^{2}/2D+1}\frac{v-1}{v}dv\\
 & = & \frac{1}{2\gamma^{2}}\int_{1}^{\gamma\rho^{2}/2D+1}1-\frac{1}{v}dv\\
 & = & \frac{1}{2\gamma^{2}}(\gamma\frac{\rho^{2}}{2D}-\log(\gamma\frac{\rho^{2}}{2D}+1)).
\end{array}
\end{equation}
Going back to C, we obtain 
\begin{equation}
\begin{array}{ccc}
C & = & \frac{\rho^{3}\arctan(\sqrt{\frac{\gamma}{2D}}\rho)}{3}-\left(2D\right)^{3/2}\frac{\sqrt{\gamma}}{3}\frac{1}{2\gamma^{2}}\left(\frac{\gamma}{2D}\rho^{2}-\log(\gamma\frac{\rho^{2}}{2D}+1)\right)\\
 & = & \frac{\rho^{3}\arctan(\sqrt{\frac{\gamma}{2D}}\rho)}{3}-\frac{\sqrt{2D}}{6\sqrt{\gamma}}\rho^{2}+\frac{\sqrt{2}D^{3/2}}{3\gamma^{3/2}}\log(\gamma\frac{\rho^{2}}{2D}+1).
\end{array}
\end{equation}

Combining all equations, using this time $C$ to denote the cell concentration, we finally obtain
\begin{equation}
\begin{array}{ccc}
K(\rho) & = & \frac{4}{3}\pi\rho^{3}+\frac{2\lambda}{C}\left(\frac{\rho^{2}}{4D}+\frac{\sqrt{\gamma}\rho^{3}\arctan(\sqrt{\frac{\gamma}{2D}}\rho)}{6\sqrt{2}D^{3/2}}-\frac{\rho^{2}}{12D}+\frac{\log\left(\gamma\frac{\rho^{2}}{2D}+1\right)}{6\gamma}-\frac{\sqrt{\gamma}\pi\rho^{3}}{12\sqrt{2}D\sqrt{D}}\right)\\
 & = & \frac{4}{3}\pi\rho^{3}+\frac{\lambda}{3C}\left(\frac{\rho^{2}}{D}+\frac{\sqrt{\gamma}\rho^{3}\arctan(\sqrt{\frac{\gamma}{2D}}\rho)}{\sqrt{2}D^{3/2}}+\frac{\log\left(\gamma\frac{\rho^{2}}{2D}+1\right)}{\gamma}-\frac{\sqrt{\gamma}\pi\rho^{3}}{2\sqrt{2}D\sqrt{D}}\right).
\end{array}\label{eq:K_intra_adv_app}
\end{equation}

Note that in the absence of advection, using Eq.~\ref{eq:pcf_noadv_bbm}, we obtain

\begin{equation}
\begin{array}{ccc}
g(r) & = & \frac{\lambda}{4\pi CDr}+1\\
\Rightarrow K'(r) & = & \frac{\lambda r}{CD}+4\text{\ensuremath{\pi r^{2}}}\\
\Leftrightarrow K(r) & = & \frac{\lambda r^{2}}{2CD}+\frac{4}{3}\pi r^{3}.
\end{array}\label{eq:K_intra_no_adv_steady_app}
\end{equation}

\paragraph{Time-dependent solution}

In the absence of advection ($U=0,\gamma=0$), we need to compute
a time-dependent solution. From eq. \ref{eq:pcf_noadv_bbm}, 
\begin{equation}
\begin{array}{ccc}
K(\rho) & = & \int_{0}^{\rho}r \frac{\lambda}{DC} \left\{ 1-\erf\left(\frac{r}{\sqrt{8Dt}}\right)\right\} +4\pi r^{2}dr\\
 & = & \frac{\lambda}{CD}\left(\frac{\rho^{2}}{2}-\int_{0}^{\rho}r\times\erf\left(\frac{r}{\sqrt{8Dt}}\right)dr\right)+\frac{4}{3}\pi\rho^{3}.
\end{array}\label{eq:start_K_intra_no_adv}
\end{equation}
We first compute the primitive for $\int_{0}^{\rho}r\times\erf\left(\frac{r}{\sqrt{8Dt}}\right)dr$.
We define $u=\frac{r}{\sqrt{8Dt}}$, $dr=\sqrt{8Dt}du$, then 
\begin{equation}
\begin{array}{ccc}
\int_{0}^{\rho}r\times\erf\left(\frac{r}{\sqrt{8Dt}}\right)dr & = & 8Dt\int_{0}^{\rho/\sqrt{8Dt}}u\times\erf\left(u\right)du.\end{array}
\end{equation}
We can integrate by parts, with $f=\erf(u)$ and $g'=u$, and obtain
\begin{equation}
\begin{array}{ccc}
8Dt\int_{0}^{\rho/\sqrt{8Dt}}u\times\erf\left(u\right)du & = & 8Dt\left(\frac{\rho^{2}}{2}\frac{1}{8Dt}\erf(\frac{\rho}{\sqrt{8Dt}})-\frac{1}{\sqrt{\pi}}\int_{0}^{\rho/\sqrt{8Dt}}u^{2}e^{-u^{2}}du\right).\end{array}\label{eq:parts}
\end{equation}
We integrate by parts again, this time with $f=u$ and $g'=ue^{-u^{2}}$,
which leads to 
\begin{equation}
\int u^{2}e^{-u^{2}}du=-\frac{ue^{-u^{2}}}{2}+\frac{1}{2}\int e^{-u^{2}}du=-\frac{ue^{-u^{2}}}{2}+\frac{\sqrt{\pi}\erf(u)}{4}.\label{eq:integrate_parts}
\end{equation}
If we use Eq. \ref{eq:integrate_parts} in Eq. \ref{eq:parts}, 
\begin{equation}
\begin{array}{ccc}
8Dt\int_{0}^{\rho/\sqrt{8Dt}}u\times\erf\left(u\right)du & = & 8Dt\left(\frac{\rho^{2}}{2}\frac{1}{8Dt}\erf(\frac{\rho}{\sqrt{8Dt}})-\frac{\erf(\frac{\rho}{\sqrt{8Dt}})}{4}+\frac{1}{2\sqrt{\pi}}\frac{\rho}{\sqrt{8Dt}}e^{-\frac{\rho^{2}}{8Dt}}\right)\\
\Leftrightarrow\int_{0}^{\rho}r\times\erf\left(\frac{r}{\sqrt{8Dt}}\right)dr & = & \frac{1}{2}\erf(\frac{\rho}{\sqrt{8Dt}})(\rho^{2}-4Dt)+\frac{\sqrt{2Dt}}{\sqrt{\pi}}\rho e^{-\frac{\rho^{2}}{8Dt}}.
\end{array}
\end{equation}
We can now compute $K(\rho)$: 
\begin{equation}
\begin{array}{ccc}
K(\rho) & = & \frac{\lambda}{CD}\left(\frac{\rho^{2}}{2}-\frac{1}{2}\erf(\frac{\rho}{\sqrt{8Dt}})(\rho^{2}-4Dt)-\frac{\sqrt{2Dt}\rho}{\sqrt{\pi}}e^{-\frac{\rho^{2}}{8Dt}}\right)+\frac{4}{3}\pi\rho^{3}.\end{array}\label{eq:end_K}
\end{equation}

\bibliographystyle{ecol_let}
\bibliography{bibliography}

\clearpage

\textbf{Supporting Information} for \textit{Local intraspecific aggregation
in phytoplankton model communities: spatial scales of occurrence and
implications for coexistence}\emph{ }by C. Picoche, W.R. Young \&
F. Barraquand.

\global\long\def\thesection{S\arabic{section}}%
 \setcounter{section}{0} 
\global\long\def\thefigure{S\arabic{figure}}%
 \setcounter{figure}{0} 
\global\long\def\theequation{S\arabic{equation}}%
 \setcounter{equation}{0}

\section{Derivation of the turbulent map \label{section:si_derivation}}

We show here how to derive a discrete-time map for turbulence from
the continuous-time formula. We consider that the velocity field $\textbf{u}^{T}=(u_{x},\,u_{y},\,u_{z})$
at position $\bx^{T}=(x,\,y,\,z)$ alternates between the three dimensions
during a period $\tau$, so that 
\begin{equation}
\textbf{u}^{T}(\bx,t)=\begin{cases}
\begin{array}{cc}
\left(U\cos(ky+\phi),0,0\right) & \text{for }n\tau\leq t<(n+\frac{1}{3})\tau\\
\left(0,U\cos(kz+\theta),0\right) & \text{for }(n+\frac{1}{3})\tau\leq t<(n+\frac{2}{3})\tau\\
\left(0,0,U\cos(kx+\psi)\right) & \text{for }(n+\frac{2}{3})\tau\leq t<(n+1)\tau.
\end{array}\end{cases}\label{eq:pierrehumber_continuous_time}
\end{equation}

The discrete-time map can be obtained by computing the displacement
over a period, between $t=n\tau$ and $t+\tau=(n+1)\tau$, with $\bx(t+\tau)=\bx(t)+\textbf{\ensuremath{\int_{n\tau}^{\left(n+1\right)\tau}}u}(\bx,t)dt$,
and knowing the initial position $\mathbf{x}(t)$. This can be solved
in three steps (eqs. \ref{eq:step1_pierrehumbert}, \ref{eq:step2_pierrehumbert}
and \ref{eq:step3_pierrehumbert}). We start with
\begin{equation}
\begin{array}{cc}
x(t+\tau/3)= & x(t)+\frac{U\tau}{3}\cos(ky(t)+\phi)\\
y(t+\tau/3)= & y(t)\\
z(t+\tau/3)= & z(t).
\end{array}\label{eq:step1_pierrehumbert}
\end{equation}
Then, 
\begin{equation}
\begin{array}{cc}
x(t+2\tau/3)= & x(t+\tau/3)\\
y(t+2\tau/3)= & y(t)+\frac{U\tau}{3}\cos(kz(t)+\theta)\\
z(t+2\tau/3)= & z(t).
\end{array}\label{eq:step2_pierrehumbert}
\end{equation}
And finally, 
\begin{equation}
\begin{array}{cc}
x(t+\tau)= & x(t+\tau/3)\\
y(t+\tau)= & y(t+2\tau/3)\\
z(t+\tau)= & z(t)+\frac{U\tau}{3}\cos(kx(t+\tau)+\psi).
\end{array}\label{eq:step3_pierrehumbert}
\end{equation}
In the third step, we need $z$ to be a function of $x(t+\tau)$,
not $x(t)$, so that the volume is conserved (the determinant of the
Jacobian matrix is equal to 1).

\section{Characteristics of standard spatial point processes \label{section:si_simple_point_processes}}

\begin{figure}[H]
\begin{centering}
\includegraphics[width=.79\textwidth]{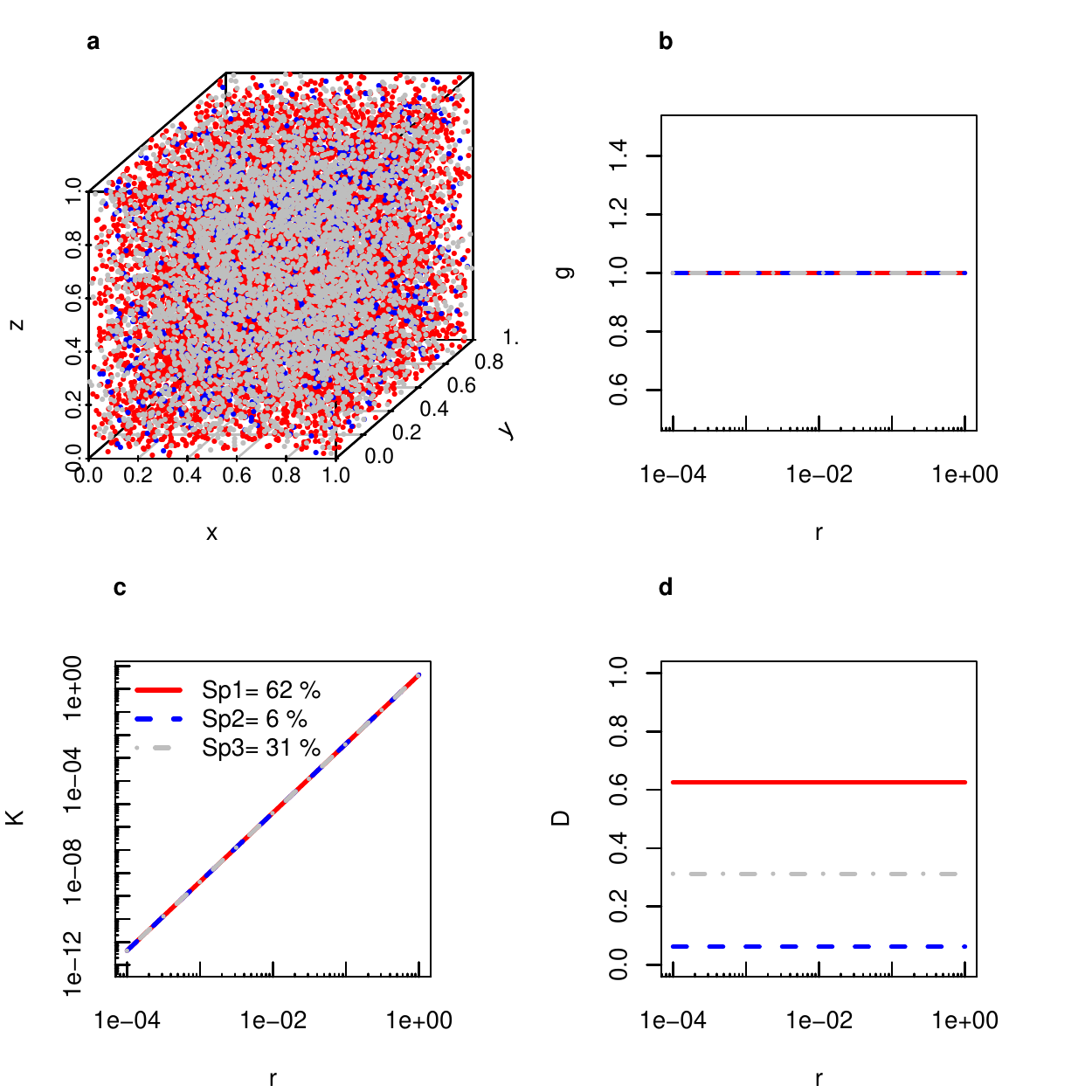}
\par\end{centering}
\caption{Example of spatial distribution (a) and theoretical pair correlation
function (b), Ripley's $K$-function (c) and dominance index (d) for
a Poisson point process in 3-species communities with different intensities
(10000 cm$^{-3}$, 1000 cm$^{-3}$, 5000 cm$^{-3}$; proportions in
the community are given in the figure). \label{fig:Example-Poisson}}
\end{figure}

\begin{figure}[H]
\begin{centering}
\includegraphics[width=0.79\textwidth]{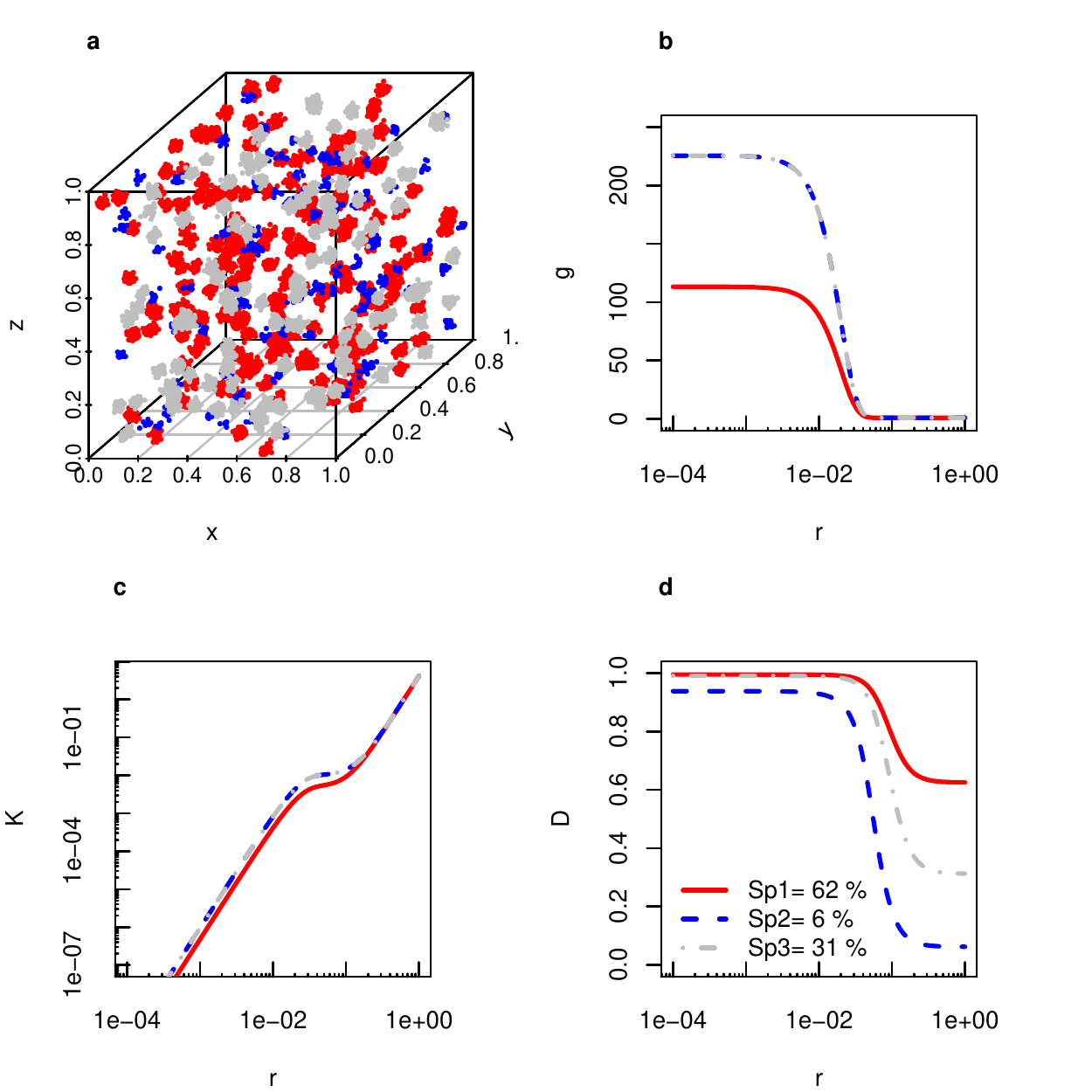}
\par\end{centering}
\caption{Example of spatial distribution (a) and theoretical pair correlation
function (b), Ripley's $K$-function (c) and dominance index (d) for
a Thomas point process in 3-species communities with different parent
intensities (200 cm$^{-3}$, 100 cm$^{-3}$, 100 cm$^{-3}$), and
different children per parent intensities (50, 10, 50; final proportions
in the communities are given in the figure), with $\sigma=0.01$.\label{fig:Example-Thomas}}
\end{figure}

In order to get the reader acquainted with the spatial point process
metrics that we use in the main text, we present here the analytical
formulas and corresponding figures (Fig. \ref{fig:Example-Poisson}
and \ref{fig:Example-Thomas}) for the pair correlation function,
Ripley's $K$-function and dominance index for well-known point processes.
We focus on the uniform distribution, i.e., the Poisson point process,
and a clustered distribution, the Thomas point process. The Thomas
point process is the result of a two-stage mechanism: a Poisson point
process generates ``parent points'' around which ``daughter points''
are scattered, their locations following a Gaussian distribution centered
on the parent location, with standard deviation $\sigma$. The numbers
of parents and daughters per parent follow two Poisson distributions
with mean $N_{p}$ and $N_{d}$ respectively. All solutions are given
for three-dimensional spatial distributions.

\subsection{Pair correlation function}

In the case of a Poisson point process, 
\begin{equation}
\forall r\geq0\text{, }g_{ii}(r)=1.
\end{equation}

For a Thomas point process, the expected value of the pcf is 
\begin{equation}
g_{ii}(r)=1+\frac{1}{C_{p}}\frac{1}{\left(4\pi\sigma^{2}\right)^{3/2}}e^{-\left(\frac{r^{2}}{4\sigma^{2}}\right)}
\end{equation}
where $C_{p}=N_{p}/V$ is the concentration/intensity of the parent
process in the volume $V$.\\

For a random superposition of stationary point processes with marks
(species) $i$ and $j$, $\forall i\neq j,\forall r\geq0\text{, }g_{ij}(r)=1$
\citepSIcite[p. 326, eq. 5.3.13]{illian2008statistical}.

\subsection{Ripley's $K$-function}

In the case of a Poisson point process, 
\begin{equation}
\forall r\geq0\text{, }K_{ii}(r)=\frac{4}{3}\pi r^{3}.\label{eq:k_poisson}
\end{equation}

For a Thomas point process, 
\begin{equation}
K_{ii}(r)=\frac{4}{3}\pi r^{3}+\frac{1}{C_{p}\sigma\sqrt{\pi}}\left(\sigma\sqrt{\pi}\erf\left(\frac{r}{2\sigma}\right)-re^{-\left(\frac{r}{2\sigma}\right)^{2}}\right).\label{eq:k_thomas}
\end{equation}

For a random superposition of stationary point processes, $K_{ij}(r)=\frac{4}{3}\pi r^{3}$
\citepSIcite[p. 324, eq. 5.3.5]{illian2008statistical}.

\subsection{Dominance index}

In the Poisson point process, $K_{ii}(r)=K_{ij}(r)$, which means
that the dominance index can be reduced to ratios of concentrations:
\begin{equation}
\mathcal{D}_{i}(r)=\frac{C_{i}}{\sum_{j=1}^{S}C_{j}}.
\end{equation}

In the Thomas process, using eq. \ref{eq:k_thomas}, 
\begin{equation}
\mathcal{D}_{i}(r)=\frac{C_{i}\left(\frac{4}{3}\pi r^{3}+\frac{F(r)}{C_{p,i}}\right)}{C_{i}\frac{F(r)}{C_{p,i}}+\sum_{j}C_{j}\frac{4}{3}\pi r^{3}}
\end{equation}
with $F(r)=\frac{1}{\sigma\sqrt{\pi}}\left(\sigma\sqrt{\pi}\erf\left(\frac{r}{2\sigma}\right)-re^{-\left(\frac{r}{2\sigma}\right)^{2}}\right)$.

\section{Convergence in time of the spatial characteristics of the BBM \label{section:si_convergence}}

\begin{figure}[H]
\begin{centering}
\includegraphics[height=0.82\textheight]{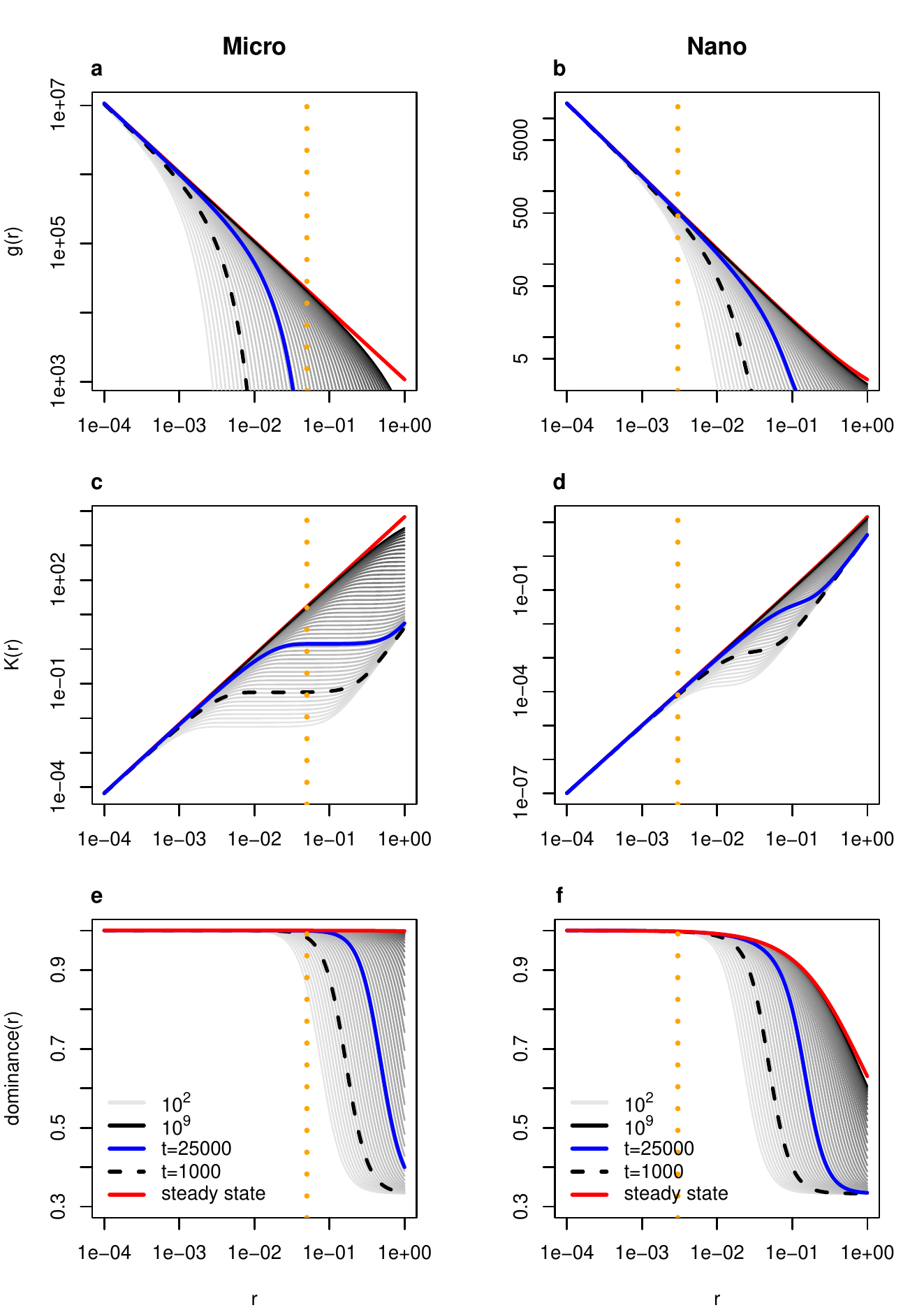} 
\par\end{centering}
\caption{Intraspecific pair correlation function (a, b), Ripley's $K$-function
(c,d) and dominance index (e,f) as a function of distance (in cm)
for microphytoplankton and nanophytoplankton in the absence of advection,
for a single species in a 3-species community with an even abundance
distribution. Shorter timeframes are shown with light grey lines while
longer ones are shown with darker shades. The theoretical value at
steady state is shown in red. The duration currently used in the simulations
($t=1000\tau$) is shown with dashed black lines. A duration corresponding
to a week is shown with solid blue lines. Dotted orange lines correspond
to the distance threshold for interaction.\label{fig:Theoretical-convergence}}
\end{figure}

The theoretical formulas of $g$, $K$ and $\mathcal{D}$ can be used
to study the behaviour of the BBM. In the absence of advection, convergence
cannot be reached in a reasonable timeframe: even a week is not long
enough for the steady-state solution to be reached (see blue line
in Fig. \ref{fig:Theoretical-convergence}). However, the population-at-equilibrium
hypothesis that we use cannot hold for such a long amount of time,
which led us to use the time-dependent formulas shown in Eqs. 5, 9
and 12 in the main text.

\begin{figure}[H]
\begin{centering}
\includegraphics[width=0.8\textwidth]{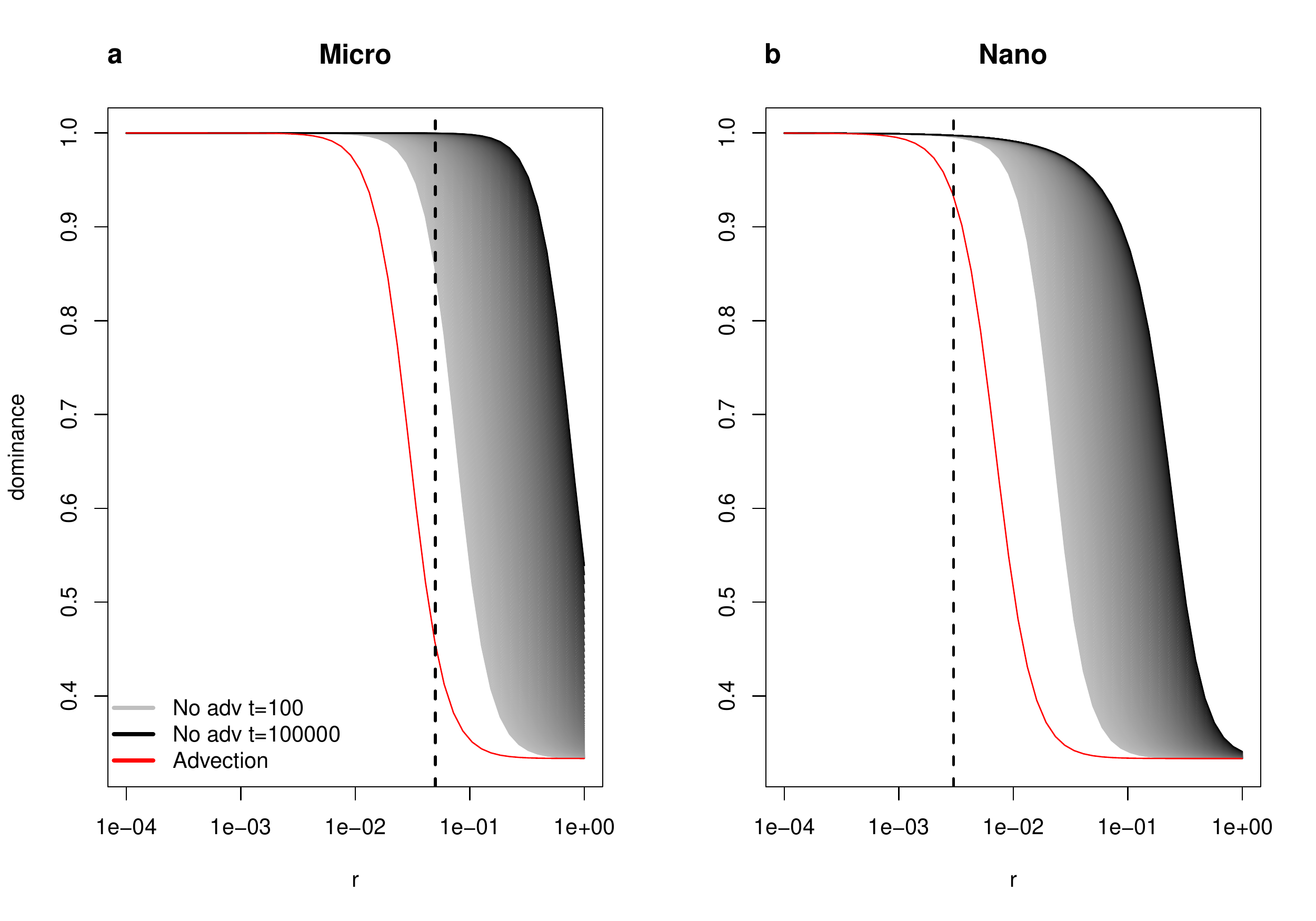} 
\par\end{centering}
\caption{Theoretical dominance indices as a function of the distance (in cm)
from a particle of a given species, for a microphytoplankton (a) and
nanophytoplankton (b) 3-species community with an even abundance distribution,
with (red line) and without (grey to black lines, with darker lines
for longer simulations) advection. The vertical dashed line corresponds
to the distance threshold for interaction. \label{fig:Theoretical-dom}}
\end{figure}

In a similar fashion, we can show with the dominance index (Fig. \ref{fig:Theoretical-dom})
the progressive clustering of individuals with time when advection
is absent, and compare it to the steady state this time \emph{with}
advection. We see that even after a short period of time ($t=100\tau$),
the dominance index without advection is larger than with advection,
and this spatial aggregation only grows with time in absence of turbulent
advection.

\section{Computation of the pair correlation function and Ripley's $K$-function\label{section:si_computation}}

The algorithm for pcf computation was adapted from the function \verb|pcf3est|
in spatstat 2.2-0 \citepSIcite{baddeley_spatstat} and slightly modified
to compute the interspecific pcf (i.e., the pcf for marked point processes).

The pcf estimate $\hat{g}_{ij}(r)$ is computed via the use of the
Epanechnikov kernel $\kappa_{E}$ with bandwidth $\delta$, i.e. 
\begin{equation}
\begin{array}{ccc}
\hat{g}_{ij}(r) & = & \frac{1}{\hat{C_{i}}}\frac{1}{\hat{C_{j}}}\frac{1}{4\pi r^{2}}\sum_{k\in i}\sum_{l\in j}\kappa_{E}(r-||\bx_{k}-\bx_{l}||)w(\bx_{k},\bx_{l})\end{array}\label{eq:pcf_estimate}
\end{equation}
where $w(\bx_{k},\bx_{l})$ is the Ohser translation correction estimator
\citepSIcite{ohser_estimators_1983} and the kernel is defined as follow.
\begin{equation}
\kappa_{E}(x)=\begin{cases}
\frac{3}{4\delta}\left(1-\frac{x^{2}}{\delta^{2}}\right) & \text{for }-\delta\leq x\leq\delta\\
0 & \text{otherwise}.
\end{cases}
\end{equation}

The estimate $\hat{g}_{ij}(r)$ is therefore very sensitive to the
bandwidth: if it is too small, the estimate is noisy and may even
be missing several pairs of points; if it is too large, the smoothing
might be so important that values are strongly underestimated. In
spatstat 2.2-0 \citepSIcite{baddeley_spatstat}, the bandwidth default value
is $\delta=0.26C^{-1/3}$. The pcf computation function was first
tested on standard distributions (with the default bandwidth), then
on the Brownian Bug Model (with different bandwidths, see Fig. \ref{fig:bandwidth_BBM}).

Estimates of the Ripley's $K$-function were also computed with the
Ohser translation correction estimator but did not require any kernel
smoothing. The same computation could be done using wrapped-around
boundary conditions (for pcf estimation; for simulation we always
consider periodic boundary conditions).

\subsection{Standard point processes}

\begin{figure}[H]
\begin{centering}
\includegraphics[height=0.77\textheight]{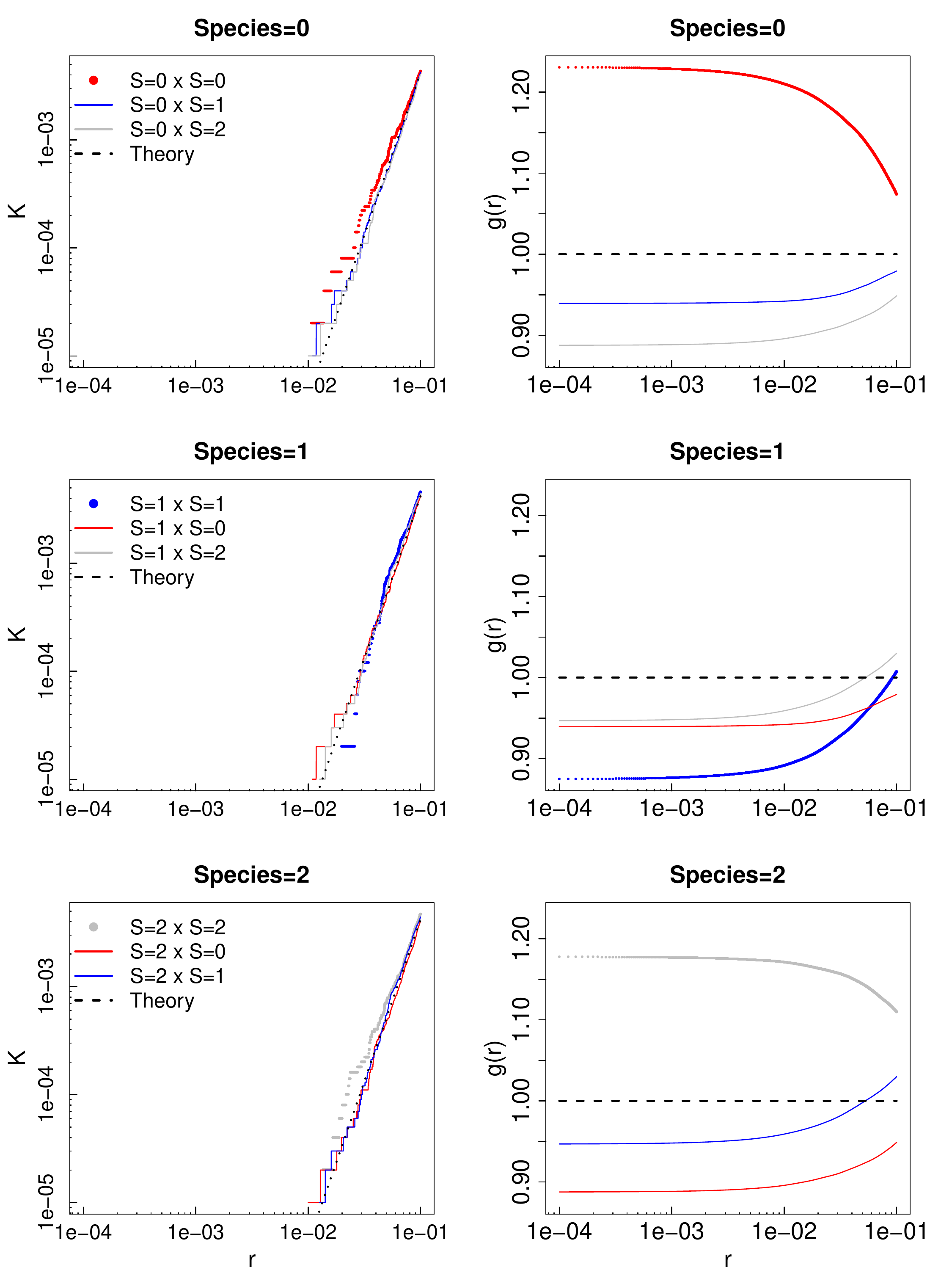} 
\par\end{centering}
\caption{Intra- and inter-specific Ripley's $K$-function and pair correlation
function values as a function of distance (in cm) for 3 species following
a Poisson process with intensity 10 cm$^{-3}$, in a volume of 1000
cm$^{3}$. Values computed from our simulations (circles and solid
lines for intra- and interspecific values, respectively) are compared
with theoretical formulas (dotted lines). Note that theoretical values
are the same for intra and interspecific indices for the Poisson distribution.
Colors correspond to the different species (red for species 0, blue
for species 1, grey for species 2).}
\end{figure}

\begin{figure}[H]
\begin{centering}
\includegraphics[height=0.77\textheight]{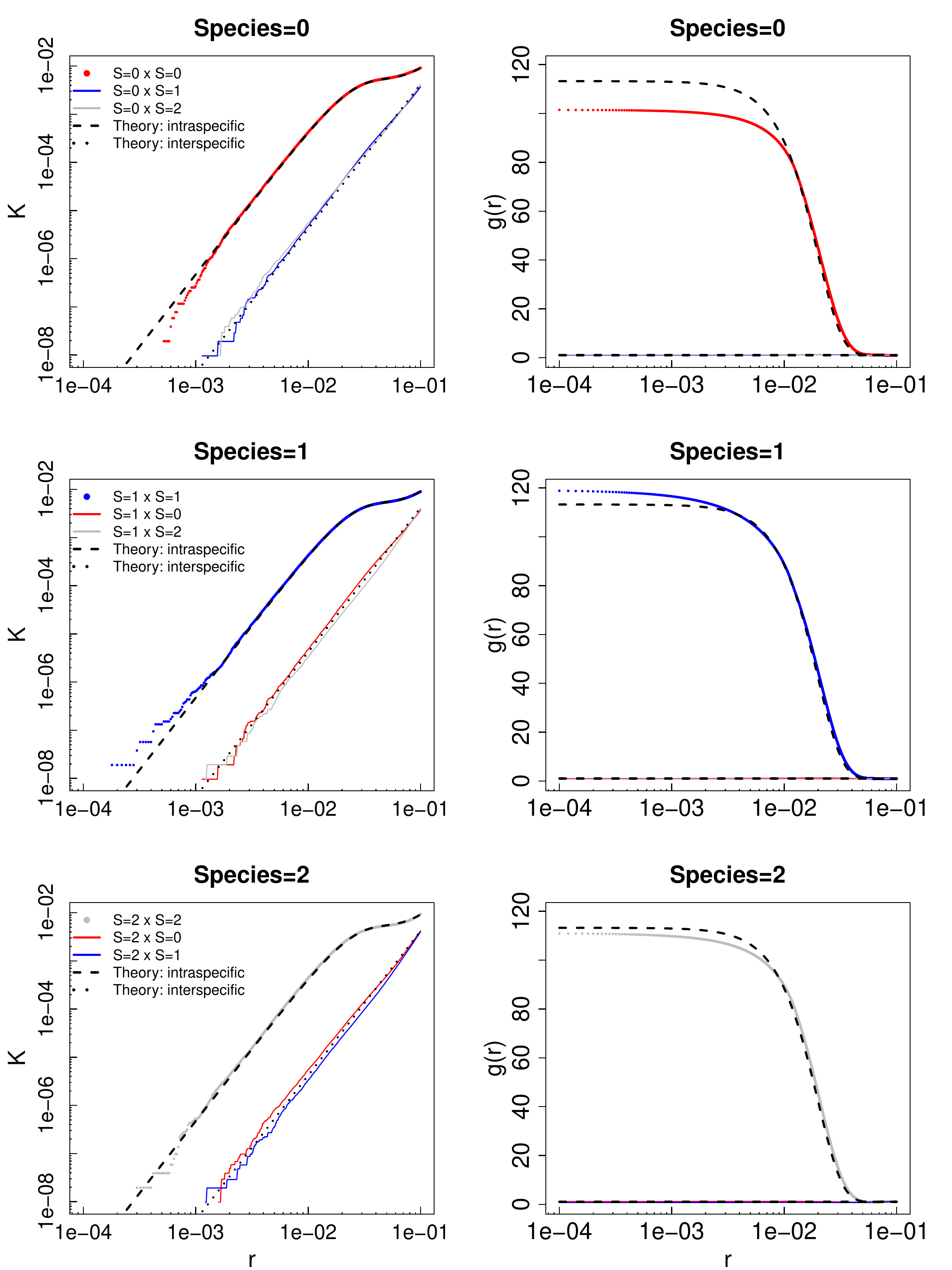} 
\par\end{centering}
\caption{Intra- and inter-specific Ripley's $K$-function and pair correlation
function values as a function of distance (in cm) for 3 species following
a Thomas process with parent intensity $C_{p}=200$ cm$^{-3}$, number
of children per parent $N_{c}=50$, in a volume of 1 cm$^{3}$, $\sigma=0.01$
and $\delta\approx0.01$2. Values computed from our simulations (circles
and solid lines for intra- and interspecific values, respectively)
are compared with theoretical formulas (dashed and dotted lines for
intra- and interspecific values, respectively). Colors correspond
to the different species (red for species 0, blue for species 1, grey
for species 2). }
\end{figure}

\subsection{Brownian Bug Model}

\begin{figure}[H]
\begin{centering}
\includegraphics[width=0.6\textwidth]{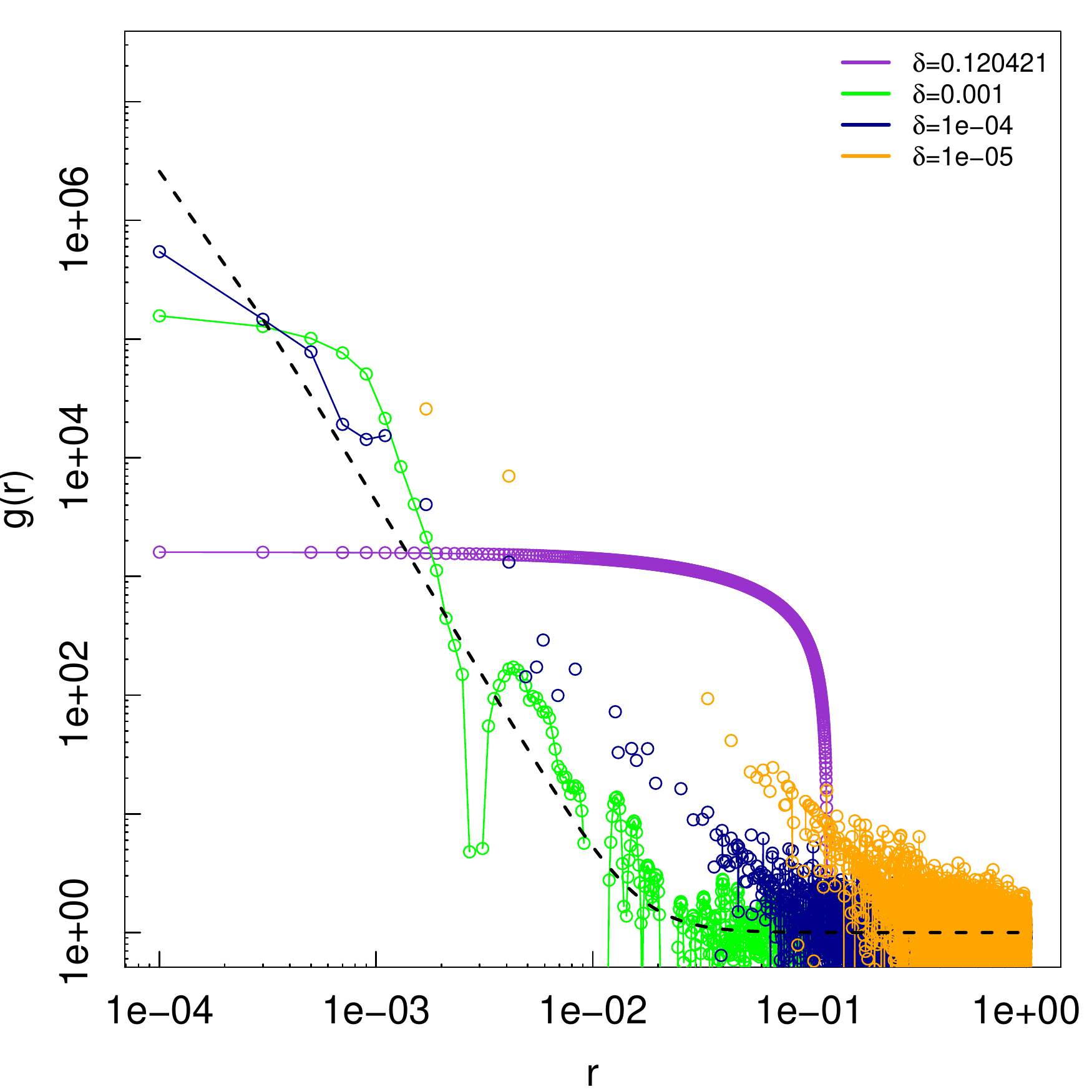} 
\par\end{centering}
\caption{Intraspecific pair correlation function as a function of distance
(in cm) computed for the Brownian Bug Model with microphytoplankton
individuals, after 1000 time steps, with different values of the bandwidth
$\delta$. The dashed line indicates the theoretical pcf.\label{fig:bandwidth_BBM} }
\end{figure}

While the pcf was one of the first indices that we intended to use,
we quickly realized that the combination of the large range of distances
we wanted to explore (from $10^{-4}$ to 1 cm) and the low density
of individuals, at least for microphytoplankton, made the estimation
difficult as the choice of the bandwidth was critical. We give an
example of the sensitivity of the pcf computation to the bandwidth
below (Fig. \ref{fig:bandwidth_BBM}).

We decided, realizing that it would be very challenging to obtain
a non-noisy pcf curve matching the theoretical expectation, to focus
on Ripley's $K$-function whose cumulative nature helps the estimation
process, which enabled us to compute the dominance index without having
to calibrate a bandwidth beforehand.

\section{Spatial distributions \label{section:si_spatial_distrib}}

\begin{figure}[H]
\begin{centering}
\includegraphics[width=0.99\textwidth]{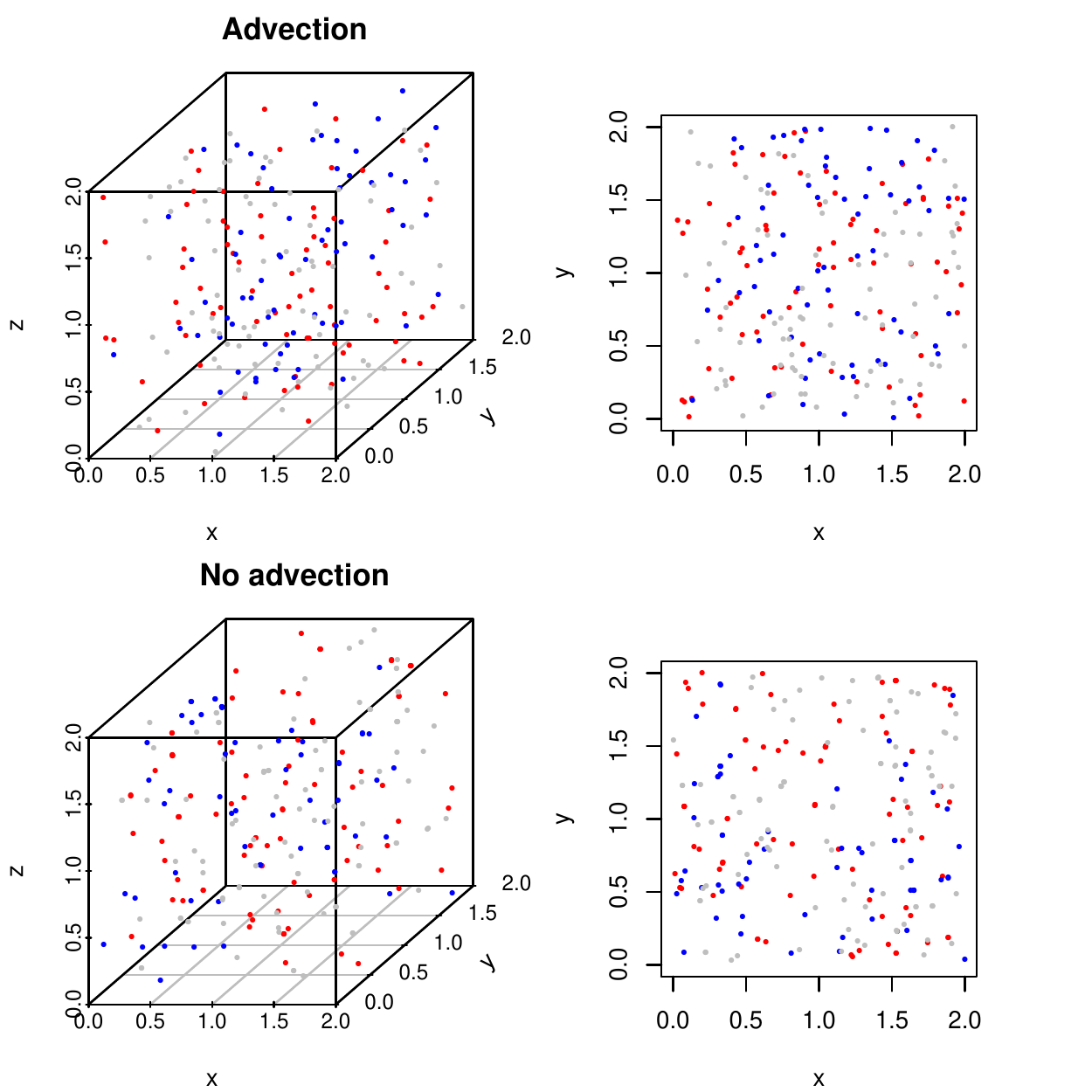} 
\par\end{centering}
\caption{Spatial distributions of a 3-species community of microphytoplankton
with and without advection with density $C=10$ cells~cm$^{-3}$
after 1000 time steps. Each color corresponds to a different species.
On the left-hand side, only a zoom on a $2\times2\times2$ cm$^{3}$
cube is shown, and its projection on the x-y plane is shown on the
right-hand side. \label{fig:Spatial-distributions-SI}}
\end{figure}

\section{Sensitivity to the computation of the advection parameter \label{section:si_advection_param}}

\begin{figure}[H]
\begin{centering}
\includegraphics[width=0.99\textwidth]{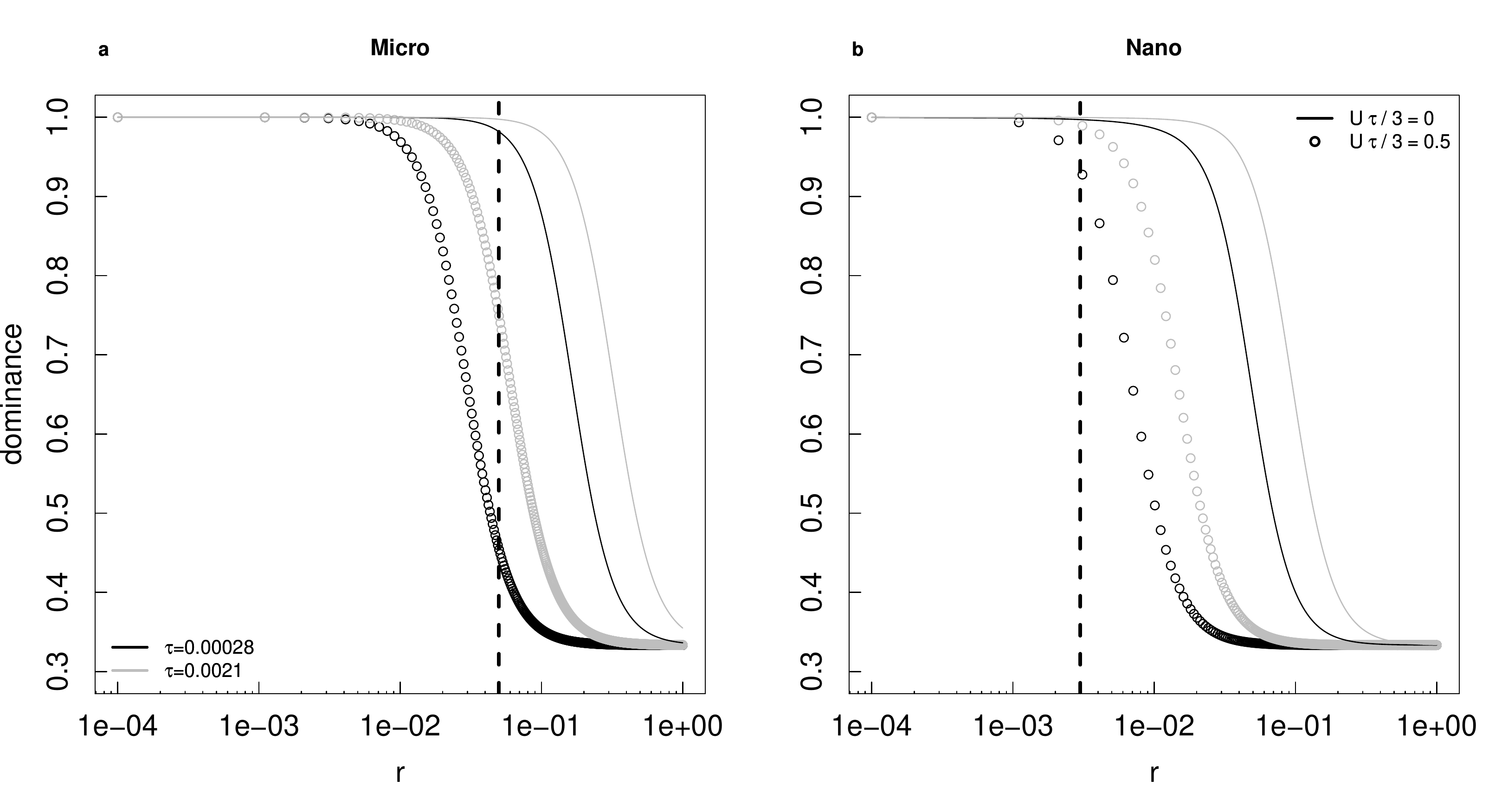} 
\par\end{centering}
\caption{Dominance indices as a function of distance (in cm) for one species
in a microphytoplankton (a) and nanophytoplankton (b) 3-species community
with even distributions after 1000 timesteps with (circles) and without
(lines) advection for different durations of the timesteps, with reference
parameters (black) and lower flow velocity (grey). \label{fig:Dominance_advection}}
\end{figure}

To compute the value of the maximum velocity of an organism in our
model at the Kolmogorov scale, we used the formula $\text{Re}=U/k\nu\approx1$
where $k$ is the smallest wavenumber associated with turbulence.
However, we could compute the Reynolds number with another, slightly
different formula, using the equivalent sphere diameter ($L_{v}$)
of our system (eq. \ref{eq:re_lv}). In this case, $\text{Re}=UL_{v}/\nu$
and 
\begin{equation}
\begin{array}{cccc}
 & \frac{4}{3}\pi\left(\frac{L_{v}}{2}\right)^{3} & = & L_{c}^{3}\\
\Leftrightarrow & L_{v} & = & 2L_{c}\left(\frac{3}{4\pi}\right)^{1/3}\\
\Leftrightarrow & L_{v} & = & 1.24\text{ cm}.
\end{array}\label{eq:re_lv}
\end{equation}
If we use $U\approx\nu/L_{v}$, we obtain $U\approx8.1\times10^{-5}$ m~s$^{-1}$. Using $U\tau/3=0.5$ cm, we then have $\tau=185\text{ s}=2.1\times10^{-3}$ d. This value of $U$ is associated to an estimated $\gamma=164$ d$^{-1}$. As could be expected,
when the flow velocity $U$ decreases, mixing decreases (Fig. \ref{fig:Dominance_advection}).

Fig. \ref{fig:Dominance_advection} also functions as a sensitivity analysis of our results with respect to the technical characterization of $U$ (resp. $\gamma$): while decreasing $U$ (resp. $\gamma$) decreases the mixing, so that microphytoplankton could in fact be slightly more aggregated, the dominance index never gets above 0.7 at the interaction radius threshold---the results are not modified substantially. However, combination of such lower $U$ and a slightly lower interaction threshold (see Discussion) may create some intraspecific dominance in microphytoplankton too. 

\section{Minimum distances between individuals \label{section:si_min_distances}}

\subsubsection*{Theory}

One of the reasons why estimating $K(r)$, and even more so $g(r)$,
is difficult is that for small distances (below 10$^{-2}$), we can
find very few observations of pairs of points. In order to better
understand at which distance ranges we should expect some estimation
difficulties, we wanted to compute the minimum expected distance between
points (distance to the nearest neighbour, DNN) when they are uniformly
distributed.

In $d$ dimensions, the probability distribution of the distance $r$
to the nearest-neighbour follows $f(r)=db_{d}Cr^{d-1}\exp(-b_{d}r^{_{d}}C)$
where $C$ is the intensity of the process. If we want to find the
distribution of the minimum DNN between $n$ realized points of a
Poisson process with intensity $C$, we can write 
\begin{equation}
\begin{array}{ccc}
\mathbb{P}(\min(R_{1},...,R_{n})>r) & = & \mathbb{P}(R_{1}>r,...,R_{n}>r)\\
 & = & \Pi_{i=1}^{n}\mathbb{P}(R_{i}>r)\\
 & = & \Pi_{i=1}^{n}\exp(-b_{d}r^{_{d}}C)\\
 & = & \exp(-b_{d}r^{_{d}}\Sigma_{i=1}^{n}C).
\end{array}
\end{equation}

We can then conclude that the distribution of the minimum distance
follows the same distribution as the DNN, but with intensity $nC$.

\medskip{}

\citetSIcite{clark_generalization_1979} show that a variable with probability
distribution (with notations changed to fit our own) $f(r)=\frac{dC\pi^{d/2}r^{d-1}}{\Gamma(\frac{d}{2}+1)}\exp(-\frac{C\pi^{d/2}r^{d}}{\Gamma(\frac{d}{2}+1)})=dCb_{d}r^{d-1}\exp(-Cb_{d}r^{d})$
has an expected value of $\mu_{d}=\frac{\left(\Gamma(\frac{d}{2}+1)\right)^{1/d}\Gamma(\frac{1}{d}+1)}{C^{1/d}\pi^{1/2}}$.

With intensity $nC$, we can write $\frac{\left(\Gamma(\frac{d}{2}+1)\right)^{1/d}\Gamma(\frac{1}{d}+1)}{(nC)^{1/d}\pi^{1/2}}$.

\medskip{}

In three dimensions, 
\begin{equation}
\begin{array}{ccc}
\mu_{d} & = & (nC)^{-1/3}\frac{\left(\Gamma(\frac{3}{2}+1)\right)^{1/3}\Gamma(\frac{1}{3}+1)}{\pi^{1/2}}\\
 & = & (nC)^{-1/3}\left(\frac{3}{2}\Gamma(3/2)\right)^{1/3}\frac{1}{3}\Gamma(1/3)\frac{1}{\pi^{1/2}}\\
 & \approx & 0.554\frac{1}{(nC)^{1/3}}.
\end{array}\label{eq:dnn_realization}
\end{equation}

This needs to be taken into account when defining $C$. For microphytoplankton,
using $C=10$ cells~cm$^{-3}$ and $n\approx10^{4}$, the expected
smallest NN distance for a uniform distribution is $1.2\times10^{-2}$
cm. For nanophytoplankton, using $C=10^{3}$ cells~cm$^{-3}$ and
$n\approx10^{4}$, it is reduced to $2.6\times10^{-3}$ cm. \medskip{}

\subsubsection*{Simulations}

\begin{figure}[H]
\begin{centering}
\includegraphics[width=0.85\textwidth]{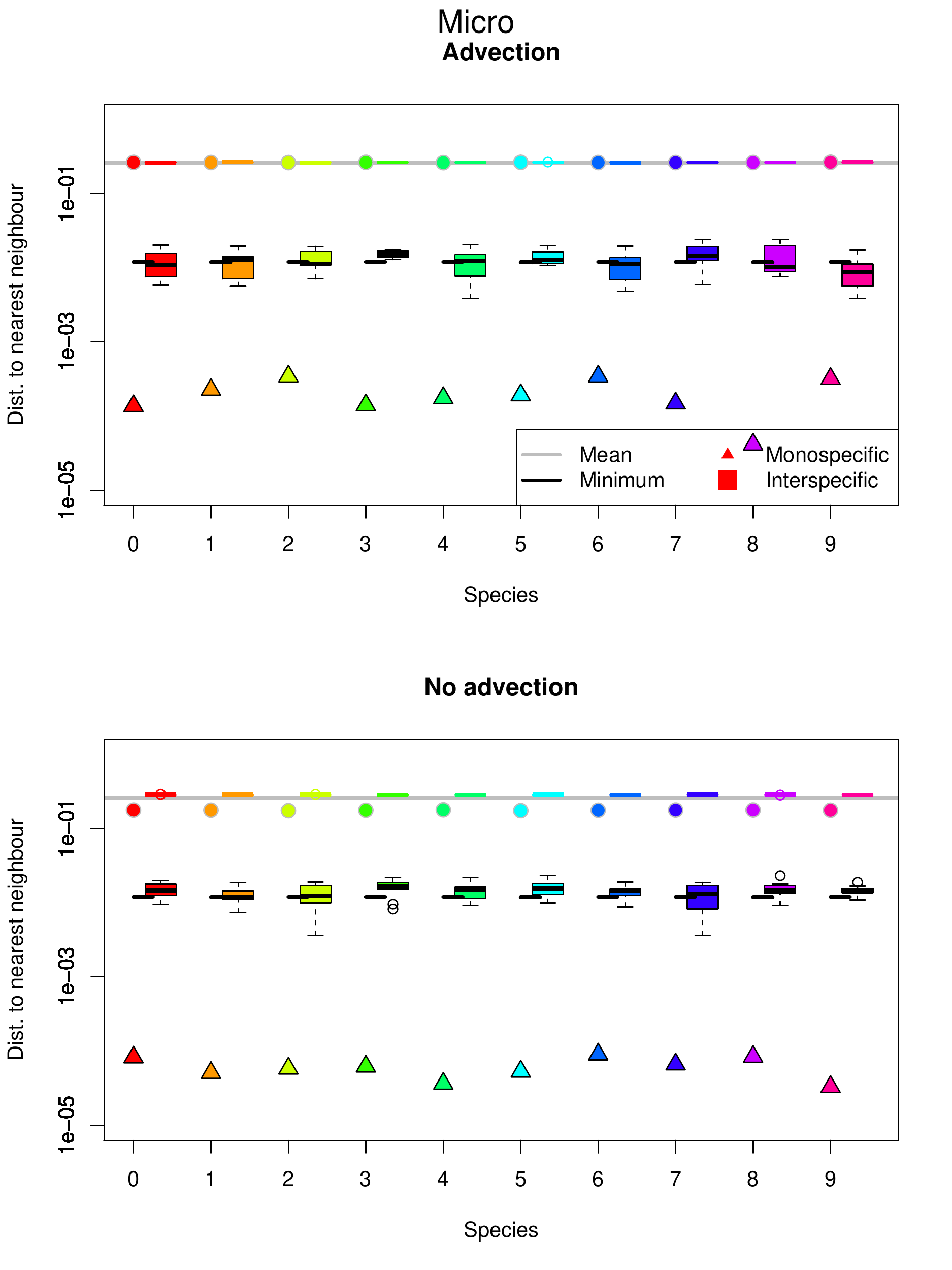} 
\par\end{centering}
\caption{Mean and minimum distance (in cm) to the nearest neighbour for 10
microphytoplankton species with density $C=10$ cells~cm$^{-3}$,
with and without advection, after 1000 time steps, compared to predictions
for a uniform distribution. Horizontal lines show the average distance
to the nearest neighbour (grey line) and the expected minimum distance
to the nearest neighbour with the actual number of realizations (black
line). Circles and triangles represent mean and minimum distance to
a conspecific, respectively. Boxplot corresponds to the distribution
of mean (grey outlines) and minimum (black outlines) distances to
a heterospecific. Colors correspond to different species. \label{fig:Distance_micro}}
\end{figure}

We can compute the simulated distance to the nearest neighbour in
the BBM and compare it to what we should obtain with a uniform spatial
distribution: the simulated mean distance to the nearest organism,
regardless of its species, is close to the expected value under a
uniform distribution, but the minimum distance to a conspecific is
much lower than expected (Fig. \ref{fig:Distance_micro} for microphytoplankton,
results are similar for nanophytoplankton).

\subsubsection*{Relationship with densities}

In the case of a uniform distribution, an increase in density leads
to a decrease in distance to the nearest neighbour (eq. \ref{eq:dnn_realization}).
Mechanically, we can indeed expect that if the number of particles
increases within the same volume, they likely get closer to each other.
We confirm that this is also the case in the Brownian Bug Model.

\begin{figure}[H]
\begin{centering}
\includegraphics[width=0.75\textwidth]{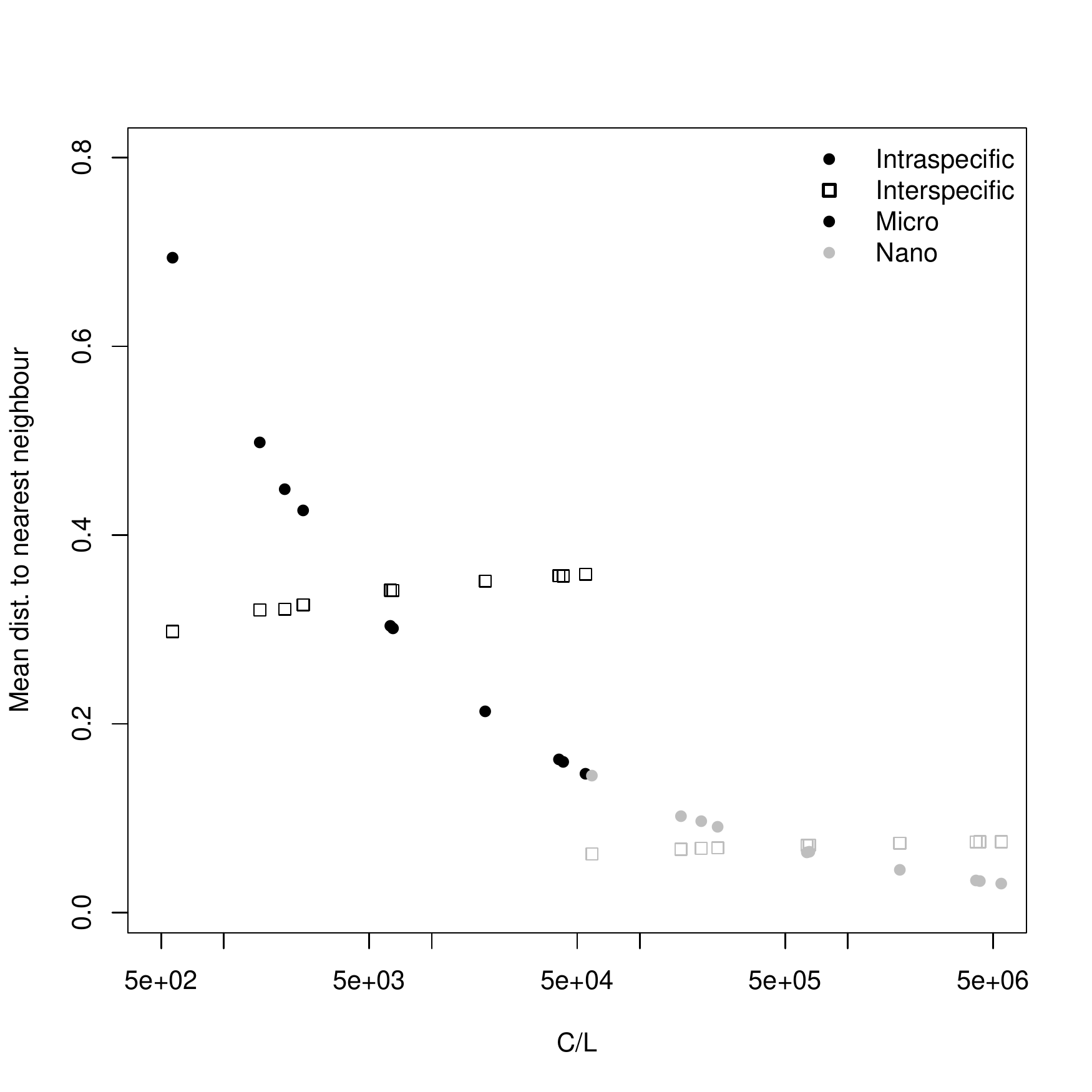} 
\par\end{centering}
\caption{Mean distance (in cm) to the nearest conspecific (filled circle) or
heterospecific (empty square) as a function of density in the environment
for both microphytoplankton (black) and nanophytoplankton (grey) communities
with a skewed abundance distribution, in the presence of advection.\label{fig:Distance_abundance}}
\end{figure}

\section{Relationship between the dominance index, relative strengths of interactions
and coexistence in Lotka-Volterra models \label{section:si_pop_interactions}}

In this section, we evaluate the potential relationship between local
dominance, ratios of intra-to-interspecific interaction strengths
observed at the population level, and their consequences in a spatial,
dynamic point process Lotka-Volterra framework. Let us define $\mu_{i}$
the average growth rate of a typical individual of species $i$. Assuming
it is linearly dependent on the abundances of the individual's conspecifics
and heterospecifics within a neighbourhood of radius $r$, 
\begin{equation}
\begin{array}{ccc}
\mu_{i}(r,t) & = & b_{i}+\beta_{ii}K_{ii}(r,t)C_{i}(t)+\beta_{io}\Sigma_{j\neq i}K_{ij}(r,t)C_{j}(t)\\
 & = & b_{i}+\beta_{ii}M_{ii}(r,t)+\beta_{io}M_{io}(r,t)
\end{array}\label{eq:lv_individual}
\end{equation}
where $b_{i}$ is the intrinsic individual growth rate, and $\beta_{ii}$/$\beta_{io}$
are individual-level interaction coefficients with a conspecific /
heterospecific, respectively. $C_{j}(t)K_{ij}(r,t)$ is the expected
number of individuals of species $j$ around a typical individual
of species $i$ within a sphere of radius $r$ centered on the focal
individual at time $t$. $M_{ii}(r,t)~=~K_{ii}(r,t)C_{i}(t)$ and
$M_{io}(r,t)~=~\Sigma_{j\neq i}K_{ij}(r,t)C_{j}(t)$.

If we are close to an equilibrium at the local scale, and intra- and
interspecific interaction strengths are equal \textit{at the individual
level} ($\beta_{ii}=\beta_{io}=\beta$), on average, 
\begin{equation}
b_{i}+\beta M_{ii}(r,t)+\beta M_{io}(r,t)\approx0.\label{eq:lv_individual_equilibrium}
\end{equation}

\medskip{}

We can now focus on the dynamics at the community level. We denote
$\alpha_{ij}$ the interactions at population level (by contrast to
$\beta_{ij}$ at individual level, as in \citealpSIcite{wiegand_consequences_2021}).
Assuming that all \emph{interspecific} population-level interactions
are similar to one another so that $\alpha_{ij}=\alpha_{io}$ if $j\neq i$,
the per capita growth rate at population level can be written as
\begin{equation}
\begin{array}{ccc}
\frac{1}{C_{i}}\frac{dC_{i}(t)}{dt} & = & b_{i}+\alpha_{ii}C_{i}(t)+\alpha_{io}C_{o}(t)\end{array}\approx0.\label{eq:lv_population}
\end{equation}

We can then write the approximate equalities $\alpha_{ii}\approx\beta\frac{M_{ii}(r,t)}{C_{i}(t)}$
and $\alpha_{io}\approx\beta\frac{M_{io}(r,t)}{C_{o}(t)}$ by matching
Eqs. \ref{eq:lv_individual} and \ref{eq:lv_population}, and obtain
the population-level interaction strength ratio 
\begin{equation}
\begin{array}{ccc}
\frac{\alpha_{io}}{\alpha_{ii}} & \approx & \frac{M_{io}(r,t)}{M_{ii}(r,t)}\frac{C_{i}(t)}{C_{o}(t)}.\end{array}
\end{equation}

Using the formula for the dominance index 
\begin{equation}
\begin{array}{ccc}
\mathcal{D}_{i}(r,t) & = & \frac{M_{ii}(r,t)}{M_{ii}(r,t)+M_{io}(r,t)}\\
\Leftrightarrow\frac{M_{io}(r,t)}{M_{ii}(r,t)} & = & \frac{\left(1-\mathcal{D}_{i}(r,t)\right)}{\mathcal{D}_{i}(r,t)}.
\end{array}\label{eq:dominance_supp}
\end{equation}
Thus the population-level interaction strength ratio can be written
out as a function of the dominance index and of the ratio of conspecific
to heterospecific density: 
\begin{equation}
\frac{\alpha_{io}}{\alpha_{ii}}\approx\frac{\left(1-\mathcal{D}_{i}(r,t)\right)}{\mathcal{D}_{i}(r,t)}\frac{C_{i}(t)}{C_{o}(t)}.
\end{equation}

Let us first focus on microphytoplankton in a 3-species community
with an even distribution of abundances. We know that $\mathcal{D}(d_{\text{threshold}})\approx0.4$
at equilibrium. In this case, $\frac{\left(1-\mathcal{D}_{i}(r,t)\right)}{\mathcal{D}_{i}(r,t)}\frac{C_{i}(t)}{C_{o}(t)}=0.75$.
For nanophytoplankton, $\mathcal{D}(d_{\text{threshold}})\approx0.9$,
thus $\frac{\alpha_{io}}{\alpha_{ii}}\approx0.06$. Both ratios of
population-level interaction strength are below 1, in spite of the
$\beta_{ii}=\beta_{io}=\beta$ assumption, and therefore meet a necessary
condition for diversity maintenance in a Lotka-Volterra model. Similar
calculations for the 10-species communities, combining small dominance
indices to low average concentrations, lead to $\frac{\alpha_{io}}{\alpha_{ii}}\ll1$,
compatible with coexistence \citepSIcite{barabas_effect_2016}.
\small
\bibliographystyleSIcite{ecol_let}
\bibliographySIcite{bibliography}
\end{document}